\documentclass{emulateapj}
\usepackage{lscape}

\newcommand{\lbol}{\mbox{$L_{bol}$}} 
\newcommand{\lsmm}{\mbox{$L_{smm}$}} 
\newcommand{\lint}{\mbox{$L_{int}$}} 
\newcommand{\lacc}{\mbox{$L_{acc}$}} 
\newcommand{\tbol}{\mbox{$T_{bol}$}} 
\newcommand{\teff}{\mbox{$T_{eff}$}} 
\newcommand{\lbolsmm}{\mbox{$L_{bol}/L_{smm}$}} 

\newcommand{\andre}{Andr\'{e}}
\newcommand{\merin}{Mer\'{i}n}

\newcommand{\bl}[1]{\mbox{\boldmath$ #1 $}}

\newcommand{\degree}{\mbox{$^{\circ}$}}

\newcommand{\um}{$\mu$m}
\newcommand{\lsun}{\mbox{L$_\odot$}}
\newcommand{\msun}{\mbox{M$_\odot$}}

\newcommand{\menv}{\mbox{$M_{core}$}}
\newcommand{\mcore}{\mbox{$M_{core}$}}
\newcommand{\mdisk}{\mbox{$M_{disk}$}}
\newcommand{\mstar}{\mbox{$M_{star}$}}
\newcommand{\renv}{\mbox{$r_{core}$}}
\newcommand{\rdisk}{\mbox{$R_{disk}$}}
\newcommand{\rstar}{\mbox{$R_{star}$}}
\newcommand{\mdotstar}{\mbox{$\dot{M}_s$}}
\newcommand{\mdotdisk}{\mbox{$\dot{M}_d$}}

\newcommand{\hcop}{HCO$^+$}

\slugcomment{\footnotesize For resubmission to ApJ, \today}

\begin{document}
\title {\bf Resolving the Luminosity Problem in Low-Mass Star 
Formation}

\author{
Michael M.~Dunham\altaffilmark{1,2} \& Eduard I.~Vorobyov\altaffilmark{3,4}
}

\altaffiltext{1}{Department of Astronomy, Yale University, P.O. Box 208101, 
New Haven, CT 06520, USA}

\altaffiltext{2}{michael.dunham@yale.edu}

\altaffiltext{3}{Institute of Astronomy, The University of Vienna, Vienna, 1180, Austria; 
eduard.vorobiev@univie.ac.at}

\altaffiltext{4}{Institute of Physics, Southern Federal University, Stachki 194, 
Rostov-on-Don, 344090, Russia.}

\begin{abstract}
We determine the observational signatures of protostellar cores by
coupling  two-dimensional radiative transfer calculations with 
numerical hydrodynamical simulations that predict accretion rates that 
both decline with time and feature short-term variability and episodic bursts caused by 
disk gravitational instability and fragmentation.  We calculate the radiative 
transfer of the collapsing cores throughout the full duration of the collapse, 
using as inputs the core, disk, and protostellar masses, radii, and mass 
accretion rates predicted by the hydrodynamical simulations.  
From the resulting spectral 
energy distributions, we calculate standard observational signatures 
(\lbol, \tbol, \lbolsmm) to directly compare to observations.  
We show that the accretion process predicted by these 
models reproduces the full spread of observed protostars in both 
\lbol\ $-$ \tbol\ and \lbol\ $-$ \mcore\ space, 
including very low luminosity objects, provides a reasonable match to the 
observed protostellar luminosity distribution, and resolves the 
long-standing luminosity problem.  These models 
predict an embedded phase duration shorter than recent observationally 
determined estimates (0.12 Myr vs.~0.44 Myr), and a fraction of total time 
spent in Stage 0 of 23\%, consistent with the range of values determined by 
observations. On average, the models spend 1.3\% of their total time in 
accretion bursts, during which 5.3\% of the final stellar mass accretes, with
maximum values being 11.8\% and 35.5\% for the total time and accreted stellar 
mass, respectively.  Time-averaged models that filter out the accretion 
variability and bursts do not provide as good of a match to the observed 
luminosity problem, suggesting that the bursts are required.
\end{abstract}



\section{Introduction}\label{sec_intro}

Low-mass stars form from the gravitational collapse of dense cores of gas and 
dust (e.g., Beichman et al.~1986; Di Francesco et al.~2007; Ward-Thompson et 
al.~2007a).  In the simplest model, the collapse of a singular isothermal 
sphere initially at rest as first considered by Shu (1977) and later extended 
by Terebey, Shu, \& Cassen (1984; TSC84) to include rotation (often called 
the ``standard model'' of star formation), collapse proceeds in an 
``inside-out'' fashion, beginning in the center of the core, moving outward at 
the sound speed, and giving rise to a constant mass accretion rate of 
$\sim 2 \times 10^{-6}$ \msun\ yr$^{-1}$.  Many modifications to this model 
have been explored, including non-zero initial inward motions (Larson 1969; 
Penston 1969; Hunter 1977; Fatuzzao, Adams, \& Myers 2004), magnetic fields 
(Galli \& Shu 1993a, 1993b; Li \& Shu 1997; Basu 1997), isothermal spheres that are 
\emph{not} singular but feature flattened density profiles at small radii 
(Foster \& Chevalier 1993; Henriksen, \andre, \& Bontemps 1997), and a finite 
outer boundary (Henriksen, \andre, \& Bontemps 1997; Vorobyov \& Basu 2005a).  
All but the latter (finite outer boundary) generally increase the accretion 
rate over that predicted by the standard model.

A significant shortcoming of the standard model is the classic ``luminosity 
problem,'' whereby accretion at the above rate produces accretion luminosities 
($L_{acc} \propto M_* \dot{M}$) factors of $10-100$ higher than typically 
observed for embedded protostars.  First noticed by Kenyon et al.~(1990) and 
further investigated by Kenyon et al.~(1994) and Kenyon \& Hartmann (1995), 
this problem has recently been emphasized by studies presenting results from 
the \emph{Spitzer Space Telescope} ``From Molecular Cores to Planet Forming 
Disks'' (cores to disks, hereafter c2d; Evans et al.~2003) Legacy Program.  
One of the first results to come from the c2d project was the discovery of 
Very Low Luminosity Objects (VeLLOs), objects embedded within dense cores with 
\lint\footnote{The internal luminosity, \lint, is the luminosity of the 
central source and excludes luminosity arising from external heating.} $\leq$ 
0.1 \lsun, most in cores previously classified as starless (Young et al.~2004; 
Bourke et al.~2006; Dunham et al.~2006, 2008, 2010a; Di Francesco et al.~2007; 
Terebey et al.~2009; Lee et al.~2009).  The discovery of such low luminosity 
protostars only exacerbated the existing luminosity problem.  Furthermore, 
Dunham et al.~(2008), Enoch et al.~(2009a), and Evans et al.~(2009) all 
examined protostellar luminosities from the c2d survey and showed that the 
protostellar luminosity distribution spans more than three orders of
 magnitude, is strongly skewed towards low luminosities (greater than 50\% of 
protostars feature luminosities indicating $\dot{M} \la 10^{-6}$ \msun\ 
yr$^{-1}$), and is inconsistent with the standard model as well as with the 
modifications described above, which tend to increase the mass accretion rate 
and thus make the problem worse.

One possible resolution to the luminosity problem is the idea that mass 
accretion is not constant.  As noted by Kenyon et al.~(1990), either accretion 
rates that decline with time or accretion rates that are very low most of 
the time but occasionally very high could resolve the luminosity problem.  
The latter process, commonly referred to as episodic accretion, features 
prolonged periods of 
lower-than-average accretion punctuated by short bursts of higher-than-average 
accretion, a scenario already invoked to explain the luminosity flares seen in 
FU Orionis objects (Hartmann \& Kenyon 1985).  First proposed by 
Kenyon et al.~(1990), such a solution was also suggested by Dunham et 
al.~(2008), Enoch et al.~(2009a), and Evans et al.~(2009) as a plausible 
explanation for both the large spread in observed luminosities and the 
significant population of sources at low luminosities.  Theoretical studies 
have provided several mechanisms for such a process in the embedded 
protostellar phase.  Hydrodynamical and MHD simulations have demonstrated 
that material accreting from a core can pile up in a circumstellar disk until 
the disk becomes gravitationally unstable and fragments into spiral structure 
and dense clumps, which are then driven onto the protostar in short-lived 
accretion bursts generated through the gravitational torques associated with 
the spiral structure (Vorobyov \& Basu 2005b, 2006, 2010; 
Machida et al.~2011).  Numerical hydrodynamic simulations without 
self-consistent disk-core interaction but with gravitationally 
overstable massive disks have also shown quick migration of dense clumps 
into the disk inner region and probably onto the star (Boss 2002, 
Cha \& Nayakshin 2011).   Alternatively, several authors have explored a 
scenario where a similar process involving gravitational instabilities in 
the outer disk triggers rapid accretion into the inner disk; the subsequent 
heating of the inner disk by this accretion activates magnetorotational 
instabilities (MRI) that then drive material onto the protostar in short 
accretion bursts (Armitage et al.~2001; Zhu et al.~2009a, 2009b, 2010).  
Other possible mechanisms have been proposed as well, including 
quasi-periodic magnetically driven outflows in the envelope causing 
magnetically controlled accretion bursts (Tassis \& Mouschovias 2005),
decay and regrowth of MRI turbulence (Simon et al.~2011), close interaction 
in binary systems or in dense stellar clusters (Bonnel \& Bastein 1992; 
Pfalzner et al.~2008), and disk-planet interaction (Lodato \& Clarke 2004;
Nayakshin \& Lodato 2011).

Direct observational evidence for episodic mass accretion bursts in the 
embedded phase is less clear.  Several Class I sources, including V1647 
Ori (e.g., \'{A}brah\'{a}m et al.~2004; Acosta-Pulido et al.~2007; Fedele et 
al.~2007; Aspin et al.~2009), OO Serpentis (K\'{o}sp\'{a}l et al.~2007), 
[CTF93]216-2 (Caratti o Garatti et al.~2011), and VSX J205126.1 (Covey et 
al.~2011; K{\'o}sp{\'a}l et al.~2011), have undergone recent flares attributed 
to accretion bursts.  However, these sources only flared in \lbol\ by factors 
of $\sim$ $2 - 10$, and all appear to be very late Class I sources near the 
end of the embedded phase with little remaining envelope material. 
On the other hand, the detection of silicate features and CO$_2$ ice bands 
in absorption in several known FU-Ori-like objects suggests the presence 
of massive envelopes (Quanz et al.~2007) and one of these objects, RHO~1B,  
has recently brightened by a factor of 1000 (Staude \& Neckel 1991).  
Episodic mass ejection is seen in jets ejected from some protostellar systems, 
suggesting an underlying variability in the mass accretion, although the 
combination of jet velocities and spacing between knots often suggest shorter 
periods of episodicity than found by the above theoretical studies (e.g., 
Lee et al.~2007; Devine et al.~2009).  Additionally, several Class 0 sources 
drive strong outflows implying higher average mass accretion rates than 
expected from their current luminosities (\andre\ et al.~1999; Dunham et 
al.~2006, 2010a; Lee et al.~2010) and Watson et al.~(2007) showed a mismatch 
between the accretion rates onto the disk and protostar of NGC 1333-IRAS 4B, 
although their results are currently under debate (Herczeg et al.~2011).  
Finally, White \& 
Hillenbrand (2004) showed that a sample of optically visible Class I sources 
in Taurus have very low accretion rates comparable to 
values observed for typical Class II sources rather than those expected for 
accreting Class I objects, although it is unclear if this result is simply a 
consequence of the Class I objects in their study being optically visible and 
thus very near the end of the embedded stage.

In this paper we test the hypothesis that the accretion process predicted by 
Vorobyov \& Basu (2005b, 2006, 2010), which includes both accretion rates 
that decline with time and episodic accretion bursts,
can resolve the luminosity 
problem and match the observed properties of embedded protostars.  We couple 
the hydrodynamic simulations presented by Vorobyov \& Basu with radiative 
transfer models to calculate the observational signatures of cores collapsing 
to form protostars in the manner predicted by these simulations, and use these 
results to directly compare to observations.  This work is a direct 
follow-up to two previous studies.  In the first, Young \& Evans (2005; 
hereafter Paper I) used a one-dimensional dust radiative transfer code to 
calculate the observational signatures of cores undergoing inside-out 
collapse following Shu (1977).  They followed three different cores with 
initial masses of 0.3, 1, and 3 \msun\ and showed that such models match 
only the upper end of the protostellar luminosity distribution, reconfirming 
the luminosity problem.  In the second study, Dunham et al.~(2010b; hereafter 
Paper II) revisited the models from Paper I with a two-dimensional radiative 
transfer code and showed that including improved dust opacities, a 
circumstellar disk and rotationally flattened envelope structure, mass-loss, 
and outflow cavities improved the match to observations but did not fully 
resolve the luminosity problem, whereas a toy-model representation 
of episodic accretion could in fact resolve the luminosity problem.  However, 
the latter conclusion is only tentative since the episodic accretion was 
included in a very 
simple manner where all mass accreted from the core builds up in the disk 
until the disk reaches 20\% of the protostellar mass, at which point the 
accretion rate from the disk onto the star jumps from 0 \msun\ yr$^{-1}$ to 
$10^{-4}$ \msun\ yr$^{-1}$ until the disk is fully drained of its mass.  The 
accretion rate from the disk onto the star then drops back down to 0 \msun\ 
yr$^{-1}$ and the cycle begins anew.  This simplification likely exaggerates 
the fraction of total mass accreted in bursts, as pointed out by Offner \& 
McKee (2011) (see also \S \ref{sec_discussion_bursts}).  Here we will revisit 
the Paper II results using accurate predictions for the evolution of 
collapsing cores from hydrodynamical simulations.

This paper is complementary to several other recent studies.  Myers et 
al.~(1998) presented simple radiative transfer calculations predicting 
observational signatures of collapsing cores, including an exponentially 
declining accretion rate with time and the effects of mass-loss, and showed 
that such models could generally reproduce the median observed protostellar 
luminosities but not the full range of values.  However, their model 
evolution is not based on a fully self-consistent analytic or numerical 
model.  Lee (2007) modified the Paper I model to include episodic accretion 
in a very simple manner in order to study the effects such a process has on 
the chemical evolution of collapsing cores.  Vorobyov (2009b) compared the 
distribution of mass accretion rates in their simulations featuring episodic 
accretion (Vorobyov \& Basu 2005b, 2006) to those inferred from the 
luminosities of protostars in the Perseus, Serpens, and Ophiuchus molecular 
clouds compiled by Enoch et al.~(2009a) and showed that their simulations 
reproduced some of the basic features of the observed distribution of mass 
accretion rates.  However, the observed distribution of accretion rates is 
quite uncertain since it is calculated from the observed protostellar 
luminosity distribution with assumed values for the protostellar masses and 
radii and with the assumption that all observed luminosity is accretion 
luminosity.  Finally, Offer \& McKee (2011) presented analytic derivations 
of the protostellar luminosity distribution for different models and 
concluded that models that tend toward a constant accretion time rather than 
constant accretion rate produce a greater spread in luminosities and are in 
better agreement with observations, similar to the result obtained by Myers 
(2010) that analytic models with accretion rates that increase with mass can 
at least partially resolve the luminosity problem.  However, both Offner \& 
McKee (2011) and Myers (2010) simply compared theoretical luminosities with 
the observed protostellar luminosity distribution, whereas this study uses 
radiative transfer calculations to ``observe'' the underlying theory (in this 
case, the Vorobyov \& Basu simulations) in a manner consistent with, and with 
direct comparison to, the observations.

The organization of this paper is as follows.  A brief description of the 
models is provided in \S \ref{sec_model}, focusing on the hydrodynamical 
simulations in \S \ref{sec_model_sim} and the radiative transfer models in 
\S \ref{sec_model_radtrans}.  \S \ref{sec_comparing} describes how the models 
are turned into observational signatures and the observational dataset to 
which the models are compared, while \S \ref{sec_results} describes the basic 
results.  A discussion of these results is presented in \S 
\ref{sec_discussion}, focusing on the degree to which the models resolve the 
luminosity problem in \S \ref{sec_discussion_lum}, the match between 
observed and model bolometric temperatures in \S \ref{sec_discussion_tbol}, 
the duration of the embedded phase in 
\S \ref{sec_discussion_durations}, and the number, duration, and importance of 
the mass accretion bursts in \S \ref{sec_discussion_bursts}.  
Finally, a basic summary of our 
results and conclusions is given in \S \ref{sec_summary}.

\section{Description of the Model}\label{sec_model}

In this section we provide a description of both the hydrodynamical 
simulations of collapsing cores used to predict the time evolution of physical 
quantities such as masses, radii, and accretion rates (\S 
\ref{sec_model_sim}), and the radiative transfer models used to calculate the 
observational signatures of collapsing cores following these simulations (\S 
\ref{sec_model_radtrans}).  The radiative transfer models are based on those 
previously presented in Papers I and II, thus we only summarize the most 
important information and refer the reader to Papers I and II for 
significantly more detailed descriptions of these models.  Additionally, we 
note here that the terms ``core'' and ``envelope'' are sometimes used 
interchangeably in the literature, whereas other times they are used 
separately to distinguish between circumstellar material of different 
densities, different states (bound vs.~unbound), etc.  In this paper we 
exclusively use the term core to refer to the full population of dense 
material which collapses to form a star, and do not use the term envelope at 
all.  In our usage the term core is interchangeable with what is commonly 
referred to as the envelope in continuum radiative transfer studies, including 
in Papers I and II.

\subsection{Hydrodynamical Simulations}\label{sec_model_sim}

\subsubsection{Basic Processes and Equations}\label{sec_model_basic}

We make use of numerical hydrodynamics simulations in 
the thin-disk approximation to compute 
the gravitational collapse of rotating, gravitationally bound, pre-stellar cores. 
This approximation is an excellent means to calculate the evolution for 
many orbital periods and many model 
realizations. It is valid as long as the aspect ratio $A$ of the disk scale 
height $h$ to radius $r$
is well below unity, which is usually fulfilled in the inner 1000 AU 
(see Figure 11 in Vorobyov \& Basu 2010).  
Protostellar disks rarely exceed 1000 AU in radius 
(Vicente \& Alves 2005; Vorobyov 2011), making 
our approach well justified for the purpose of collecting a wide statistical 
sample of model disks. At larger radial distances, $A$ may approach unity but 
those regions are not important dynamically and serve as a reservoir for 
material falling onto the disk during the embedded phase of star formation.

We start our numerical integration in the pre-stellar phase characterized by a 
collapsing {\it starless} core, 
continue into the embedded phase of star formation during which
a star and disk are formed, and terminate our simulations in the T Tauri phase
when most of the core has accreted onto the forming star+disk system.
The thin-disk approximation allows us to consider spatial scales of 
order 10000 AU and 
compute the evolution of both the core and the star+disk system altogether.
An important consequence is that the mass accretion rate onto 
the disk (\mdotdisk) is not a free parameter of our model 
but is self-consistently determined by the gas dynamics in the collapsing core.

We introduce a ``sink cell'' at $R_{\rm sc}=6$~AU and allow for the matter
in the computational domain to freely flow into the sink cell but not out of it.
We monitor the gas density in the sink cell and 
when its value exceeds a critical value for the transition from 
isothermal to adiabatic evolution ($\sim 10^{11}$~cm$^{-3}$), we introduce a 
central gravitating 
point-mass star. In the subsequent evolution, most of the gas 
that crosses the inner 
boundary is assumed to land onto the central star while a small fraction is 
assumed to be 
carried away with protostellar jets, with the exact partition being a free 
parameter (usually 10\% is assumed to be ejected). 

The main physical processes that are taken into account in our modeling include 
stellar irradiation, background irradiation with temperature $T_{\rm bg}=10$~K, 
viscous and shock heating, radiative cooling from the
disk surface, and also disk self-gravity. 
 The corresponding equations of mass, momentum, and energy transport are

\begin{equation}
\label{cont}
\hskip -5 cm \frac{{\partial \Sigma }}{{\partial t}} =  - \nabla_p  \cdot 
\left( \Sigma \bl{v}_p \right),  
\end{equation}
\begin{eqnarray}
\label{mom}
\frac{\partial}{\partial t} \left( \Sigma \bl{v}_p \right) &+& 
\left[ \nabla \cdot \left( \Sigma \bl{v_p}
\otimes \bl{v}_p \right) \right]_p =   - \nabla_p {\cal P}  + \Sigma \, 
\bl{g}_p \\ \nonumber
& + & (\nabla \cdot \mathbf{\Pi})_p, 
\label{energ}
\end{eqnarray}
\begin{equation}
\frac{\partial e}{\partial t} +\nabla_p \cdot \left( e \bl{v}_p \right) = 
-{\cal P} 
(\nabla_p \cdot \bl{v}_{p}) -\Lambda +\Gamma + 
\left(\nabla \bl{v}\right)_{pp^\prime}:\Pi_{pp^\prime}, 
\end{equation}
where subscripts $p$ and $p^\prime$ refer to the planar components $(r,\phi)$ 
in polar coordinates, $\Sigma$ is the mass surface density, $e$ is the 
internal energy per surface area, 
${\cal P}=\int^{h}_{-h} P dh$ is the vertically integrated
form of the gas pressure $P$, $h$ is the radially and azimuthally varying 
vertical scale height determined in each computational cell using an 
assumption of local hydrostatic equilibrium, 
$\bl{v}_{p}=v_r \hat{\bl r}+ v_\phi \hat{\bl \phi}$ is the velocity in the
disk plane, $\bl{g}_{p}=g_r \hat{\bl r} +g_\phi \hat{\bl \phi}$ is the 
gravitational acceleration in the disk plane, and 
$\nabla_p=\hat{\bl r} \partial / \partial r + \hat{\bl \phi} r^{-1} 
\partial / \partial \phi $ is the gradient along the planar coordinates of the 
disk.  Viscosity enters the basic equations via the viscous stress tensor 
$\mathbf{\Pi}$ and the magnitude of kinematic viscosity $\nu$ is parameterized 
using the usual $\alpha$-prescription of Shakura \& Sunyaev (1973). In our 
models, we use a spatially and temporally uniform $\alpha$, with its value 
set to $5\times 10^{-3}$.

Apart from viscous heating determined by the viscous stress tensor $\bl\Pi$, 
the thermal balance of the disk is also controlled by 
heating due to stellar and background irradiation and
radiative cooling from the disk surface. The latter is calculated using the 
diffusion approximation of the vertical radiation transport in a one-zone 
model of the vertical disk structure (Johnson \& Gammie 2003):  
\begin{equation}
\Lambda={\cal F}_{\rm c}\sigma\, T^4 \frac{\tau}{1+\tau^2},
\end{equation}
where $\sigma$ is the Stefan-Boltzmann constant, $T$ is the midplane 
temperature of gas, 
and ${\cal F}_{\rm c}=2+20\tan^{-1}(\tau)/(3\pi)$ is a function that 
secures a correct transition between the optically thick and thin regimes.  
We use frequency-integrated opacities $\tau$ of Bell \& Lin (1994).

The heating function is expressed as
\begin{equation}
\Gamma={\cal F}_{\rm c}\sigma\, T_{\rm irr}^4 \frac{\tau}{1+\tau^2},
\end{equation}
where $T_{\rm irr}$ is the irradiation temperature at the disk surface 
determined by the stellar and background black-body irradiation as
\begin{equation}
T_{\rm irr}^4=T_{\rm bg}^4+\frac{F_{\rm irr}(r)}{\sigma},
\label{fluxCS}
\end{equation}
where $T_{\rm bg}$ is the uniform background temperature 
(in our model set to the initial temperature of the natal cloud core)
and $F_{\rm irr}(r)$ is the radiation flux (energy per unit time per unit 
surface area) absorbed by the disk surface at radial distance 
$r$ from the central star. The latter quantity is calculated as 
\begin{equation}
F_{\rm irr}(r)= \frac{L_\ast}{4\pi r^2} \cos{\gamma_{\rm irr}},
\end{equation}
where the incidence angle  of radiation $\gamma_{\rm irr}$ 
arriving at the disk surface at radial distance $r$ is calculated using 
the model's known radial profile of the disk scale height $h$, and 
$L_\ast$ is the sum of the accretion luminosity $L_{\rm \ast,accr}$ arising from 
the gravitational energy of accreted gas and 
the photospheric luminosity $L_{\rm \ast,ph}$ due to gravitational compression 
and deuterium burning in the star interior. While the former quantity is 
calculated from the model's known stellar mass and accretion rate onto the 
star, the latter is taken from the pre-main-sequence tracks of 
D'Antona \& Mazzitelli (1994).  A more detailed explanation 
of the model can be found in Vorobyov \& Basu (2010).

\subsubsection{Initial Conditions}\label{sec_model_initial}

The form of the initial gas density, temperature and angular velocity profiles 
in gravitationally bound, dense pre-stellar cores may vary depending on the 
environment and physical processes that contribute to the formation of 
such cores. In this work, we consider two possible gas surface density 
$\Sigma$ and angular velocity $\Omega$ profiles and assume that the initial 
gas temperature in the core is equal to $T_{\rm init}\equiv T_{\rm bg}=10$~K. 
Most models cores have $\Sigma$ and $\Omega$ typical for magnetically 
supercritical cores formed by slow gravitational contraction (Basu 1997) 
\begin{equation}
\Sigma={r_0 \Sigma_0 \over \sqrt{r^2+r_0^2}}\:,
\label{dens}
\end{equation}
\begin{equation}
\Omega=2\Omega_0 \left( {r_0\over r}\right)^2 
\left[\sqrt{1+\left({r\over r_0}\right)^2
} -1\right],
\label{omega}
\end{equation}
where $\Omega_0$ is the central angular velocity and 
$r_0$ is the radius of central near-constant-density plateau defined 
as $r_0 = \sqrt{A} c_{\rm s}^2 /(\pi G\Sigma_0)$, where $c_{\rm s}$ is the sound 
speed.  These initial profiles are characterized by the 
important dimensionless free parameter $(\Omega_0 r_0/c_{\rm s})^2$ and have 
the property that the ratio of rotational to gravitational energy 
$\beta\approx 0.91 (\Omega_0 r_0/c_{\rm s})^2$ (see Basu 1997). 
We note that the above form of the column 
density is very similar to the integrated 
column density of a Bonnor-Ebert sphere with a positive density enhancement 
$A$ (Dapp \& Basu 2009).  
Throughout the paper, we use $A=1.2$. 
In addition, every core is truncated so that the ratio of the core 
radius to the radius of the central flat region $r_{\rm core}/r_0$ 
is constant and equal to 6.0.
The truncation mechanisms could be erosion of the core outer regions by UV 
radiation and/or tidal stripping.

\begin{deluxetable*}{lcccccc}
\tabletypesize{\scriptsize}
\tablewidth{0pt}
\tablecaption{\label{tab_models}Model Parameters}  
\tablehead{
\colhead{} & \colhead{$\Omega_0$} & \colhead{$r_0$} & \colhead{$\Sigma_0$} & \colhead{Core Outer Radius} & \colhead{Initial Core Mass} & \colhead{} \\ 
\colhead{Model} & \colhead{(rad s$^{-1}$)} & \colhead{(AU)} & \colhead{(g~cm$^{-2}$)} & \colhead{(pc)} & \colhead{(\msun)}  & \colhead{$\beta$} 
}
\startdata
1 & $1.6 \times 10^{-13}$ &685 & 0.18 & 0.02 & 0.305 & $8.75\times10^{-3}$ \\
2 & $1.95 \times 10^{-13}$ &685 & 0.18 & 0.02 & 0.305 & $1.26\times10^{-2}$ \\
3 & $3.0 \times 10^{-14}$ & 3770 & 0.033 & 0.11 & 1.684 & $8.8\times10^{-3}$ \\
4\tablenotemark{a} & $2.5 \times 10^{-14}$ & -- & 0.026  & 0.04 & 0.612 & $8.75\times10^{-3}$ \\
5\tablenotemark{a} & $0.9 \times 10^{-14}$ & -- & $9.3\times10^{-3}$ & 0.11 & 1.686 &$8.8\times 10^{-3}$  \\
6\tablenotemark{a} & $1.2 \times 10^{-14}$ & -- &0.013 & 0.08 & 1.22 & $8.8\times 10^{-3}$ \\
7 & $2.0 \times 10^{-13}$ & 445 & 0.28 & 0.013 & 0.194 & $5.6\times10^{-3}$ \\
8 & $2.3 \times 10^{-14}$ & 3770 & 0.033 & 0.11 & 1.689 & $5.6\times 10^{-3}$ \\
9 & $2.6 \times 10^{-13}$ & 514 & 0.24 &  0.015 & 0.229 & $1.26\times10^{-2}$ \\
10 & $2.9 \times 10^{-14}$ & 3085 & 0.04 & 0.09 & 1.378 & $5.6\times 10^{-3}$ \\
11 & $3.2 \times 10^{-14}$ & 4115 & 0.03 & 0.12 & 1.84 & $1.26\times10^{-2}$ \\
12 & $3.7 \times 10^{-14}$ & 2400 & 0.05 &0.07 & 1.0726 & $5.6\times 10^{-3}$ \\
13 & $3.8 \times 10^{-14}$ & 1645 & 0.075 & 0.048 & 0.7337 & $2.75\times 10^{-3}$\\
14 & $3.9 \times 10^{-13}$ & 342 & 0.36 & 0.01 & 0.1515 & $1.26\times10^{-2}$ \\
15 & $3.9 \times 10^{-14}$ & 3430 & 0.036 & 0.1 & 1.5312 & $1.26\times10^{-2}$ \\
16 & $3.25 \times 10^{-13}$ & 274 & 0.45 & 0.008 & 0.1174 & $5.6\times 10^{-3}$\\
17 & $4.7 \times 10^{-14}$ & 1885 & 0.066 & 0.055 & 0.8434 & $5.6\times 10^{-3}$ \\
18 & $4.8 \times 10^{-14}$ & 2745 & 0.045 & 0.08 & 1.2422 & $1.26\times10^{-2}$ \\
19 & $5.56 \times 10^{-13}$ & 240 & 0.52 & 0.007 & 0.1052 &  $1.26\times10^{-2}$ \\
20 & $1.1 \times 10^{-13}$ & 1200 & 0.1 & 0.035 & 0.5349 &  $1.26\times10^{-2}$\\
21 & $6.0 \times 10^{-14}$ & 1370 & 0.09 & 0.04 & 0.6084 & $5.6\times 10^{-3}$\\
22 & $2.8 \times 10^{-14}$ & 2230 & 0.056 & 0.065 & 0.999 & $2.75\times 10^{-3}$\\
23 & $2.0 \times 10^{-14}$ & 2915 & 0.043 & 0.085 & 1.304 & $2.75\times 10^{-3}$
\enddata
\tablenotetext{a}{Constant surface density profile model.}
\end{deluxetable*}

The actual procedure for generating a specific core is as follows. First, 
we fix $\beta$ between approximately $10^{-3} - 10^{-2}$ based on observational results that find that $\beta$ ranges between approximately $10^{-4} - 10^{-1}$ (Goodman et al.~1993; Caselli et al.~2002)\footnote{While both Goodman et al.~(1993) and Caselli et al.~(2002) find that approximately 50\% of observed cores have $\beta$ between $10^{-2} - 10^{-1}$, we are unable to consider models with $\beta \ga 10^{-2}$ due to technical limitations and thus may be biased against cores with high initial rotation and angular momentum.}.  
Then we fix the core radius 
$r_{\rm core}$ and find $r_0$ from the condition $r_{\rm core}/r_0=6$.  
The central surface density $\Sigma_0$ is found from the relation 
$r_0=\sqrt{A}c_{\rm s}^2/(\pi G \Sigma_0)$ and the resulting core mass 
$M_{\rm core}$ is determineed from Equation~(\ref{dens}). Finally, the central 
angular velocity $\Omega_0$ is found from the condition 
$\beta=0.9(\Omega_0 r_0/c_{\rm s})^2$.

As an alternative to the $\Sigma\propto r^{-1}$ and $\Omega\propto r^{-1}$ 
profiles, we also consider several models with radially constant $\Sigma$ 
and $\Omega$ distributions, 
corresponding to self-gravitating, sheetlike cores with volume density 
depending only on distance from the midplane 
$\rho(z)=\rho(0) \mathrm{sech}^2(z/h)$ (Boss \& Hartmann 2001). 

\begin{figure}
\epsscale{1.0}
\plotone{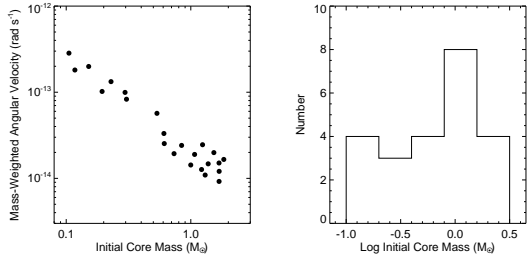}
\caption{\label{fig_masshist}\emph{Left:} Mass-weighted initial angular 
velocity for the 23 models considered in this paper and listed in 
Table \ref{tab_models}.  \emph{Right:} Histogram showing the distribution 
of initial core masses in log space with bins 0.3 dex wide.}
\end{figure}

In total, we have considered 23 models with initial core masses (\mcore) 
ranging from 0.105~$M_\odot$ to 1.84~$M_\odot$ and initial ratios of 
rotational to gravitational energy $\beta=(0.275-1.26)\times 10^{-2}$.  
The parameters of these 23 models are summarized in Table~\ref{tab_models}.  
The left panel of Figure \ref{fig_masshist} plots the mass-weighted initial angular 
velocity vs.~initial core mass for each model, where we have chosen to plot 
the mass-weighted rather than central angular velocity for direct comparison 
to observations.  Our models range from about $10^{-14}$ rad s$^{-1}$ to a few 
times $10^{-13}$ rad s$^{-1}$, consistent with the observed range found 
by Goodman et al.~(1993).  The right panel of Figure \ref{fig_masshist} plots 
the distribution of initial core masses for each model.
The absence of models with \mcore\ $>2.0~M_\odot$ is 
caused by numerical difficulties associated with modeling the collapse of
massive cores. We plan to extend our parameter space in a future study.  
This distribution clearly does not follow the stellar initial 
mass funciton (IMF).
However, all results in this paper are presented after 
weighting each individual model by the IMF in order to properly compare to 
observations (see \S \ref{sec_results} for details on this weighting).  
Our results are thus robust to the exact number of models at different masses 
as long as we adequately sample all relavant mass ranges.

\begin{figure*}
\epsscale{0.75}
\plotone{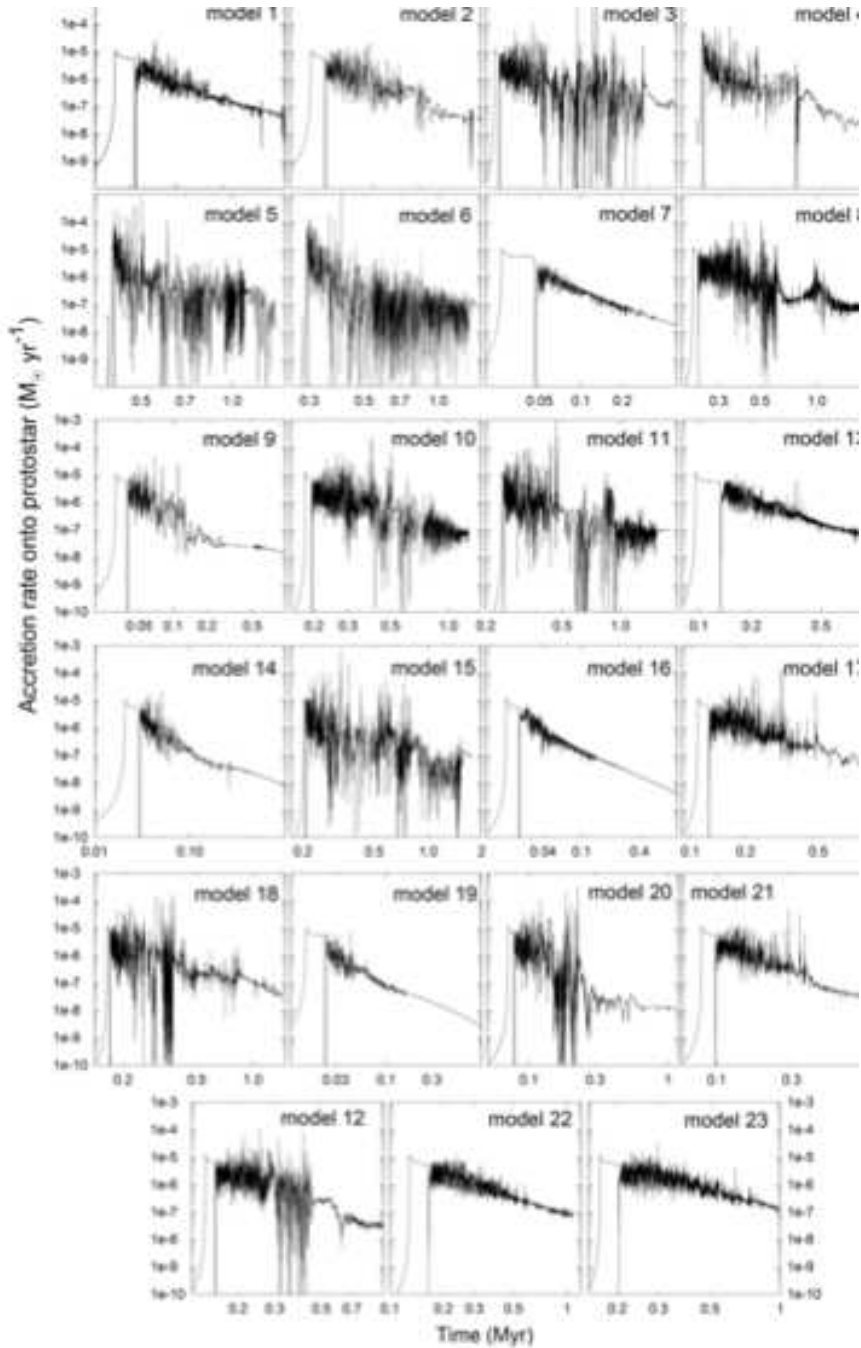}
\caption{\label{fig_mdot}Time evolution of the mass accretion rate onto 
the protostar for each of the 23 models considered in this paper and listed 
in Table \ref{tab_models}.  The zero time is defined from the beginning of the collapse of pre-stellar
cores.  Note that the full duration of each model is shown here, whereas 
the embedded phase is considered to end when 10\% of the initial core mass 
remains (see \S \ref{sec_model_radtrans}).  
Note that the 
scale of the x axis varies from one panel to the next depending on the 
total duration of each model.}
\end{figure*}

Figure \ref{fig_mdot} plots the time evolution of the accretion rate 
onto the protostar (\mdotstar) for each of the 23 models considered in this 
paper and listed in Table \ref{tab_models}.  The accretion is highly 
variable with episodic bursts of rapid accretion, with the exact amplitude 
and frequency of the variability differing from one model to the next due 
to the different initial conditions (in particular initial core mass; see 
Vorobyov \& Basu 2010 for details).  The origin of the accretion 
variability seen in Figure \ref{fig_mdot} is discussed in detail below in 
\S \ref{sec_simulations_accretion}.
One particular feature worth noting in Figure~\ref{fig_mdot} is that
the average accretion rates show a gradual decline with time. On timescales of
$\sim$~1.0~Myr the decline in $\dot{M}$ may reach three orders of magnitude. 
This behaviour is likely caused by a gradual decline of the disk mass with time,
which is particularly prominent starting from the late Class I phase when
mass loading from the parent core diminishes (Vorobyov 2009a, 2011).

\subsubsection{Disk Gravitational Instability and Variable Accretion}
\label{sec_simulations_accretion}

As was discussed in \S \ref{sec_intro}, episodic accretion is a plausible
solution to the luminosity problem and various physical processes have been 
invoked to produce episodic and repetitive bursts of mass accretion onto the 
star.  Among them, disk gravitational instability and fragmentation in the 
embedded phase of star formation has been shown to trigger intense bursts of 
accretion luminosity as the forming fragments are torqued into the disk inner 
regions and onto the star (Vorobyov \& Basu 2005, 2006, 2010; 
Machida et al.~2010). Apart from these FU-Ori-like bursts, gravitationally 
unstable disks are generally characterized by highly variable accretion onto 
the star caused by the non-linear interaction of dominant spiral modes 
(Vorobyov 2009b) or by the presence of a massive planet (Machida et al.~2010).  
In this section, we take model 12 as a prototype model and use this model to 
illustrate the role of disk gravitational instability and fragmentation
in driving variable accretion onto the star.

\begin{figure}[hbt!]
\epsscale{1.0}
\plotone{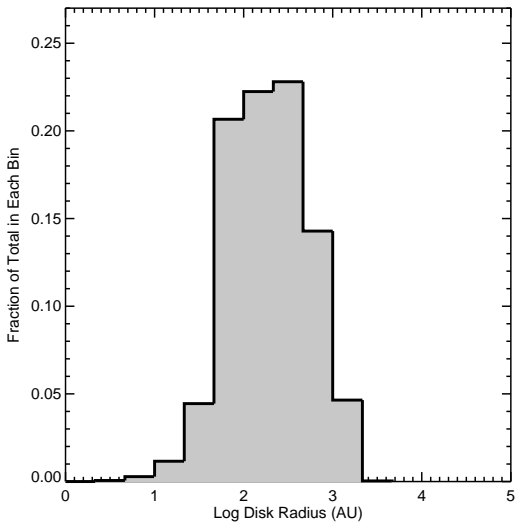}
\caption{\label{fig_rdisk}Histogram showing the fraction of total time 
spent by all models at various \rdisk, weighted by the initial mass 
function as described in \S \ref{sec_results}.  The binsize is 1/3 dex.}
\end{figure}

First, we must discuss the method by which we calculate basic disk 
properties.  The disk mass and radius are calculated at 
each timestep by disentangling 
the disk and infalling core on the computational mesh.  We do this in 
practice by adopting a surface density threshold of 0.5 g cm$^{-2}$ between 
the disk and core and also using the radial gas velocity profile.  
This method is described in detail by Vorobyov (2011), 
although we note that Vorobyov (2011) adopted a lower surface density 
threshold of 0.1 g cm$^{-2}$.  The values of the disk radius and mass both 
depend on the adopted threshold.  Figure \ref{fig_rdisk} plots a histogram 
showing the fraction of total time the 23 models considered in this paper spend 
at various disk radii (weighted by the initial mass function; see 
\S \ref{sec_results}).  Our choice of a threshold of 0.5 g cm$^{-2}$ is 
motivated by the fact that the resulting distribution of disk radii peaks 
between $10^2 - 10^3$ AU; adopting the original 0.1 g cm$^{-2}$ threshold 
would shift the entire distribution up by about a factor of $2-3$.  Sizes 
of {\it embedded} disks are very poorly constrained 
by observations, and are typically assumed to be on the order of 100 AU or 
less based on simple centrifugal radius arguments.  In reality, however, 
angular momentum transport will cause disks to spread to sizes greater 
than the centrifugal radii.  Furthermore, there are some limited observations 
supporting the existence of large protostellar disks (i.e., the 1 \msun, 300 
AU disk surorunding Serpens FIRS 1; Enoch et al.~2009b).  Ultimately, the 
correct threshold to adopt in order to disentagle the disk and core in 
the simulations will remain uncertain until the masses and sizes of 
protostellar disks are better constrained.

\begin{figure}[hbt!]
\epsscale{1.0}
\plotone{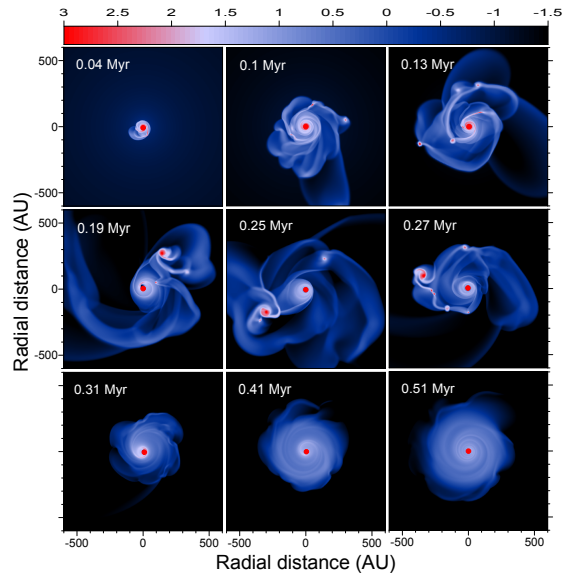}
\caption{\label{fig3}Model 12 gas surface densities in the inner 1200~AU at 
several times after the formation of the central. The star is shown in the 
coordinate center by a red circle and the time is shown 
in each panel in Myr. The scale bar is in g~cm$^{-2}$. Vigorous gravitational 
instability and fragmentation is evident in the early disk evolution at 
$0.1~\mathrm{Myr}\la t\la0.3$~Myr.}
\end{figure}

\begin{figure}[hbt!]
\epsscale{1.0}
\plotone{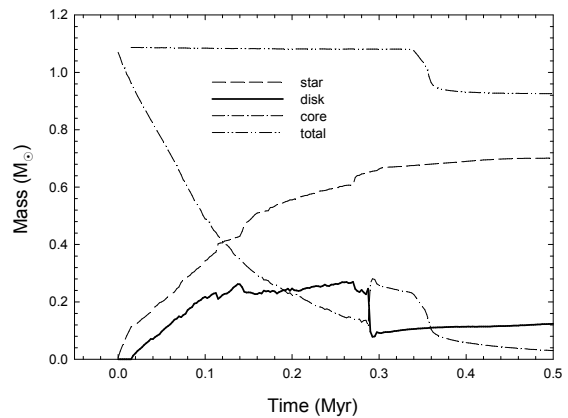}
\caption{\label{fig_masses}Time evolution of the disk, stellar and core 
masses, as well as the total mass in the computational
domain, for model~12.  The zero time is defined as the moment of formation of 
the central star.}
\label{fig6}
\end{figure}

Figure~\ref{fig3} presents gas surface density maps for model~12 at several
times after the formation of the central star. Only the inner 1200 AU are 
shown, while the 
whole computational region amounts to 28000~AU. The evolutionary times are chosen so as to
emphasize three distinct 
stages of the disk evolution, each depicted by a separate row of images.  
The upper row shows the initial stage of the disk 
formation and growth. The disk begins to form at $t\approx0.015$~Myr and 
several distinct fragment are present as early as at $t=0.1$~Myr.   The number 
of fragments varies with time due to the continuing process of disk 
fragmentation and migration of the fragments onto the star. At around 
$t=0.16$~Myr the second stage of the disk evolution begins when a massive 
fragment ($\approx 0.15~M_\odot$) forms in the disk. The fragment is prominent 
until $t=0.3$~Myr when it is ejected from the disk via many-body 
interaction with other newly-born fragments.  
This ejection event is very transient and is not captured in Figure~\ref{fig3}
but is evident in Fig.~\ref{fig6} showing the time evolution of the total mass
in our computational domain.
This is a universal mechanism that seems to occur in massive disks
and is reported in other numerical studies of disk fragmentation 
(Stamatellos \& Whitworth 2009; Bate 2009; Basu \& Vorobyov 2011). 
After the ejection event, 
the last stage of disk evolution ensues, which is shown in the bottom row of 
Fig.~\ref{fig3} and is characterized by a dramatic change in 
the disk appearance---the disk is no longer prone to fragmentation and shows only 
a weak spiral structure that diminishes with time.  We note that these three 
stages are only indicative and reflect the overall tendency of a protostellar 
disk to go through the initially vigorously unstable phase toward a marginally 
stable configuration. All three stages are often observed in relatively 
massive disks.  In low-mass disks, the first and second stages may be very short or even 
absent and the ejection event may not be present even in massive enough disks 
(Basu \& Vorobyov 2011). 

The three stages of the disk evolution are reflected in the mass accretion 
rate onto the star, $\dot{M_{\rm s}} = - 2\pi R_{\rm sc} v_r \Sigma$, 
which we calculate as the mass passing through the sink cell per one time step 
of numerical integration (which in physical units is usually equal to 10--20 
days and the total integration time can go beyond 1~Myr). The latter value is 
also corrected for the jet efficiency to account for the mass that pass through the sink cell but is
later evacuated with the jets. We 
note that the adopted size of the sink cell $R_{\rm sc} = 6$~AU is larger than 
the stellar radius, except for the very early stage when the forming star is 
represented by the first hydrostatic core. 
Decreasing $R_{\rm sc}$ entails a significant increase in the calculation times
because the physical size of the computational zones in the $\phi$-direction 
decreases as one approaches the singularity at $r$=0, causing a correponding 
decrease in the time step.  Simultaneously, the number of grid zones in the 
$r$-direction needs to be increased to sustain an adequate resolution at 
$r\sim 100$~AU on the log-spaced grid. This effectively limits our choice of 
$R_{\rm sc}$ to a few AU at best considering the available computational 
resources.  We acknowledge that our results may be sensitive to this 
limitation.  For instance, additional accretion bursts may be triggered in the 
inner disk at $r < 6$~AU due to the thermal ionization instability (Bell \& 
Lin 1994), MRI (Zhu et al.~2009a, 2009b) or disk-planet interaction (Nayakshin 
\& Lodato 2011).  We have run a few models with $R_{\rm sc}=2$~AU  and found 
that the general behaviour of the accretion rates (variability, burst 
magnitudes) remains qualitatively similar but further work is needed to extend 
the computational region still closer to the stellar surface.

\begin{figure}
\epsscale{1.0}
\plotone{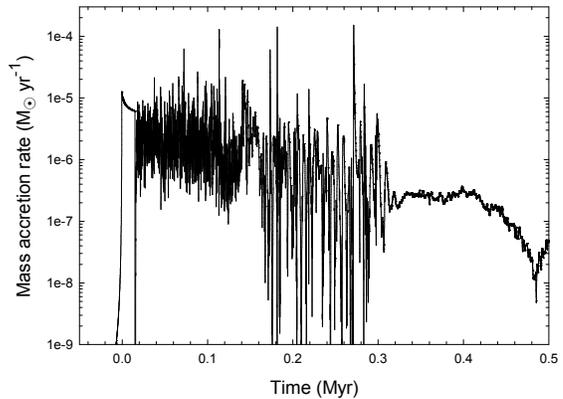}
\caption{Time evolution of the mass accretion rate onto the star for model~12.  
The zero time is defined as the moment of formation of the central star.}
\label{fig4}
\end{figure}

The time evolution of the mass accretion rate is shown in Figure \ref{fig4}. 
\mdotstar\ is negligible in the pre-stellar phase and rises to a 
maximum value of $1.3\times
10^{-5}~M_\odot$~yr$^{-1}$
when the first hydrostatic core forms at $t=0$~Myr (the time is counted from this moment). 
The subsequent short period of evolution is
characterized by a gradually declining \mdotstar\ when the material from 
the core lands
directly onto the forming star. A sharp drop in the accretion rate at 
$t=0.015$~Myr manifests the beginning
of the disk formation phase when the infalling core material hits the centrifugal 
barrier near the sink cell
and the accretion rates drops to zero. However, the process of mass loading 
from the core continues,
the disk grows in mass and size, and gravitational instability is soon ignited 
in the disk at $t=0.04$~Myr. From this
moment and until $t=0.16$~Myr, i.e., during the first stage of disk evolution, 
accretion exhibits variability with periods in the 400--1000~yr range and rates
between a $\mathrm{few} \times 10^{-7}~M_\odot$~yr$^{-1}$ and a 
$\mathrm{few} \times 10^{-5}~M_\odot$~yr$^{-1}$ and also several strong bursts with 
\mdotstar\ $\ga 2\times10^{-5}~M_\odot$~yr$^{-1}$.
The latter are caused by fragments migrating onto the star due to the loss of 
angular momentum via 
gravitational interaction with spiral arms in the disk (Vorobyov \& Basu 2005, 
2006, 2010).
The typical period of accretion variability implies that its driving force lies in the
disk intermediate regions at $r=30-60$~AU (for an average stellar mass  
$0.3~M_\odot$). Many theoretical and numerical studies, including our 
own, indicate that this is 
the minimum distance from the star at which gravitational instability and 
fragmentation can occur 
(e.g. Stamatellos \& Whitworth 2008; Clarke 2009; Vorobyov \& Basu 2010, Meru 
\& Bate 2010), suggesting
a causal link between the disk gravitational instability and accretion 
variability in our model.

During $t=0.16-0.3~$Myr, i.e., during the second stage of disk evolution,
the accretion pattern changes notably 
and starts to show a higher-amplitude and
longer-period variability, with accretion rates lying between 
$10^{-10}~M_\odot$~yr$^{-1}$ and
$\mathrm{a~few} \times 10^{-4}~M_\odot$~yr$^{-1}$. This change is caused by the
formation of a massive fragment clearly visible in the middle row in Fig.~\ref{fig3}. 
The typical period of accretion variability is now $\approx 5\times10^3$~yr, 
implying that the driving force is located at 
$r\approx 250$~AU. 
This value is an average orbital distance of the fragment, suggesting a causal link
between accretion variations and orbital motion of the fragment. 
A similar correlation between the orbital period of a massive object in the 
disk and the accretion
rate variability was found by Machida et al.~(2011).
These results demonstrate that the disk gravitational instability and 
fragmentation
are the driving forces for the accretion variability in our models.

During $t=0.3-0.5$~Myr, i.e., during the third stage of disk evolution, 
accretion onto the star again undergoes a dramatic change. There are no more
bursts or high-amplitude variations. \mdotstar\ shows a 
gradual decline with time from $\mathrm{a~few} \times 10^{-7}~M_\odot$~yr$^{-1}$ 
to $\mathrm{a~few} \times 10^{-8}~M_\odot$~yr$^{-1}$ with only low-amplitude 
flickering. 
This change is caused by the ejection of the massive fragment from the disk 
and the associated loss of a significant fraction of the disk
material. As a result, the disk settles into a state with only marginal gravitational 
instability characterized by a week and diffuse spiral structure. 
At this stage, the accretion rate is mostly determined by 
viscous torques rather than by gravitational ones (Vorobyov \& Basu 2009a).

We caution that accretion rates may be sensitive to
our choice of the $\alpha$-parameter, which determines the magnitude of viscous
torques in our model. In the present study, we use a spatially and temporally uniform
$\alpha=5\times 10^{-3}$, a value chosen for consistency with the work of 
Vorobyov \& Basu (2009b, 2010).  These authors 
studied numerically the secular evolution of viscous and self-gravitating
disks, with particular emphasis on accretion rates, and found that 
$\alpha < 10^{-2}$ is needed to reproduce both the FU Ori accretion bursts
and the range and magnitude of accretion rates found for brown dwarfs 
and T Tauri stars.

\begin{figure}[hbt!]
\epsscale{1.0}
\plotone{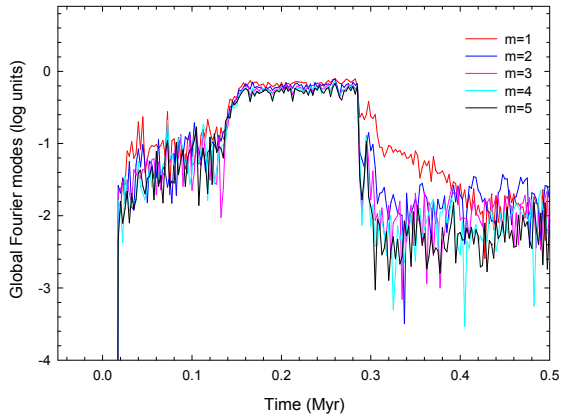}
\caption{\label{fig_fourier}Time evolution of the first five global Fourier 
amplitudes for model~12. The amplitudes are plotted in log units.  
The zero time is defined as the moment of formation of the central star.}
\end{figure}

However, the $\alpha$-parameter may vary in space and time.
In particular, if the gas surface density exceeds some critical value 
($\mathrm{a~few} \times 100$~g~cm$^{-2}$, Armitage 2011), 
then the ionization fraction due to cosmic rays and X-rays becomes insufficient to sustain 
the magneto-rotational instability and the corresponding $\alpha$-parameter is expected 
to decrease significantly. In this case, the outburst phenomenon, in particular,
and the accretion variability, in general, will be more pronounced as demonstrated 
by Vorobyov \& Basu (2009a, 2010). In our models,
surface densities in excess of the critical value are found in the inner 10--20~AU for 
relatively massive disks (Vorobyov 2010). This means that our choice of 
a fixed $\alpha=5\times10^{-3}$ may cause the outburst magnitude to damp somewhat 
in the inner parts of the disk.

Figure \ref{fig_fourier}, which plots the time evolution of the global 
Fourier modes, substantiates our analysis. 
We calculate the first five modes using the following equation
\begin{equation}
C_{\rm m} (t) = {1 \over M_{\rm disk}} \left| \int_0^{2 \pi} 
\int_{R_{\rm sc}}^{R_{\rm disc}} 
\Sigma(r,\phi,t) \, e^{im\phi} r \, dr\,  d\phi \right|,
\label{fourier}
\end{equation}
where $M_{\rm disk}$ is the disk mass and $R_{\rm disk}$ is the disk's physical 
outer radius.  

Three distinct patterns of modal behavior are seen in Figure \ref{fig_fourier}, each 
corresponding
to a particular stage of disk evolution. In the early evolution at 
$t=0.04-0.16$~Myr, the non-linear interaction between competing spiral modes 
in the gravitationally
unstable disk gives rise to
the mode variability resembling in magnitude and frequency that of the mass 
accretion rate\footnote{The output frequency with which the modes are 
calculated is lower
than that of the mass accretion rate. This should be taken into account when 
comparing Figures~\ref{fig4} and \ref{fig_fourier}}. At $t=0.16-0.3$~Myr, 
global modes are maximal (with lower order modes being 
marginally dominant) but show
little variability, while the corresponding accretion rates are highly 
variable. The matter that is
the dominant source of perturbation is now a massive compact fragment, and the 
accretion variability is
driven by the dynamics of the fragment rather than by the non-linear 
interaction of the spiral modes.  After $t=0.3$~Myr, the global modes show 
a fast decline by two orders of magnitude caused by a sharp decrease in the 
disk mass after the ejection
event. Although the modes return to a highly variable pattern, their amplitudes
are very low, $\sim 10^{-2}$, meaning that the 
non-axisymmetric density perturbations in the disk at this stage are of the order of
1\% and the mass accretion is now entirely controlled by viscous torques.
The latter drive the disk toward an axisymmetric state (Vorobyov \& Basu 2009a) and 
the corresponding accretion rates show only low-amplitude flickering, gradually declining 
as the disk is drained of its mass reservoir.

The time evolution of the disk, core and stellar mass as well as the 
total star+disk+core mass in the computational domain were shown in 
Figure \ref{fig_masses}.  It is worth noting 
that the stellar mass is always greater than that of the disk. This is a 
direct consequence of the
regulating nature of the accretion burst phenomenon. The interplay between 
mass loading from the
core and migration of the fragments onto the star helps to sustain the system 
near the fragmentation
 boundary, with the disk periodically going above and below this boundary 
until the infalling 
core is depleted of matter (Vorobyov \& Basu 2006, 2010). 
The ejection event is well visible in the time evolution of the disk, core and 
total 
mass. Both the disk and core exhibit a sharp change in mass at 
$t\approx 0.3$~Myr, 
with the disk losing mass and the core gaining mass as the fragment leaves 
the disk and 
propagates radially outward through the core. At $t\approx0.34$~Myr, the 
fragment passes through 
the outer computational boundary at r=14000~AU, which is manifested by a sharp 
drop in the total
mass in the computational domain. From the amplitude of this drop, we can 
estimate the upper limit
on the mass of the fragment as $0.15~M_\odot$. This estimate includes the 
fragment itself as well as a mini-disk that surrounds it.

\subsection{Radiative Transfer Models}\label{sec_model_radtrans}

An ideal coupling of hydrodynamic simulations with radiative transfer
models involves taking the detailed disk and core structure from the former
and using them as inputs to the latter. Our hydro simulations, however,
provide the radial structure and the disk vertical scale height
but lack the detailed vertical
density and temperature distributions (due to the adopted thin-disk
approximation) needed to run radiative transfer models.  
One way to circumvent this difficulty would be to reconstruct the disk and
core vertical structure solving for the combined equations of vertical 
hydrostatic equilibrium and vertical radiation transfer after updating the 
flow variables in the plane of the disk. This approach, which would render
our hydro simulations essentially two plus one dimensional, is currently
under development and will be presented in a future paper.

In this work, we have employed a simpler approach by taking analytic profiles
for the core and disk structure and re-scaling them according to
simulation's known parameters. More specifically, for each of
the models listed in Table \ref{tab_models}, the hydro simulations
described above provide the initial core mass (\menv) and
radius\footnote{Following the convention adopted in both Papers I and II,
radii pertaining to the core are denoted by lowercase $r$, while radii
pertaining to either the star or disk are denoted by uppercase $R$.} (\renv),
along with the time evolution of the core mass, disk mass (\mdisk),
protostellar mass (\mstar), disk outer radius (\rdisk), accretion rate onto
the protostar (\mdotstar), and accretion rate onto the disk (\mdotdisk).
These parameters are used to construct the core and disk structure
according to analytic profiles described in more detail below.  
Incorporating the exact 
physical structure at each timestep from the simulations will not 
fundamentally alter our results or conclusions since the protostellar, 
disk, and core masses, radii, and accretion rates are unchanged, but it will 
affect the detailed comparison to observations since the radiative transfer 
(and thus resuling core temperature profile, spectral energy distributions, 
and observational signatures) depend on the two-dimensional physical 
structure.  

As in Papers I and II, the core inner radius is held fixed at a value such that 
the initial optical depth at 100 \um\ is set equal to 10 (see Paper I, in 
particular Equation 4) until the disk outer radius exceeds this value; once 
this occurs the core inner radius is set equal to the disk outer radius.  
Following both Papers I and II and Vorobyov \& Basu (2010), the protostellar 
radius (\rstar) is calculated from \mstar\ following Palla \& Stahler (1991) 
(see Eq.~11 of Vorobyov \& Basu 2010), except at early times where it is 
modified to take into account the first hydrostatic core (FHSC; Larson 1969) 
stage.  
Following Vorobyov \& Basu, \rstar\ is fixed at 5~AU for the duration of the 
FHSC phase 
and then smoothly joined to the Palla \& Stahler values over a period of 
1000~yr when the second core forms.  
We use a simple polytropic model for the forming star
and assume that the transition from the FHSC to the second core (protostar)
occurs when the central temperature exceeds $2\times 10^3$~K.  This is an 
improved implementation of \rstar\ as compared to that in Papers I and 
II but it yields similar lifetimes of the FHSC, $(1-2)\times10^{4}$~yr.

As in Paper II, we adopt the core density profile given by the TSC84 solution 
for the collapse of a slowly rotating core to include the effects of 
rotational flattening.  The TSC84 solution results in a core that is initially 
a spherically symmetric, singular isothermal sphere with a density 
distribution $n \propto r_{core}^{-2}$, identical to the Shu (1977) solution.  
As collapse proceeds, the solution takes on two forms: an outer solution that 
is similar to the non-rotating, spherically symmetric solution and an inner 
solution that exhibits flattening of the density profile.  Since material 
falling in to the central regions originates from larger radii and thus 
carries more angular momentum as time progresses, the radius where the inner 
solution must be used, and thus the radius at which flattening becomes 
significant, increases with time ($r_{flat} \propto \Omega_0^2 t^3$; TSC84).

The TSC84 solution is parameterized by the initial angular velocity of the 
core and the time since the formation of the protostar.  We truncate the 
solution at 
the given \renv\ for each model and then renormalize the density profile so 
that the core mass 
matches the current \menv\ given by the hydro simulations.  As in 
Paper II, we use the velocity profiles given by the TSC84 solution to allow 
\renv\ to decrease once the infall radius\footnote{The infall radius is the 
radius within which the core is collapsing.  It starts at the center and moves 
outward at the sound speed.} exceeds the initial outer radius.  
By renormalizing the density profile 
to match the current \menv\ given by the hydro simulations our model is not 
completely physically self-consistent.  However, this choice preserves both 
the exact evolution of core mass with time given by the hydro simulations and 
the qualitative feature of the TSC84 solution that the effects of rotational flattening 
on the core density profile increase with both time and initial 
core rotation.

Following Paper II, the assumed disk structure follows a power law in the 
radial coordinate and a Gaussian in the vertical coordinate, with the density 
profile given by:
\begin{equation}\label{eq_disk_density_profile}
\rho_{disk}(s,z) = \rho_0 \left(\frac{s}{s_o} \right)^{-\alpha} 
exp\left[ -\frac{1}{2} \left(\frac{z}{H_s}\right)^2 \right] \qquad ,
\end{equation}
where $z$ is the distance above the midplane ($z=rcos\theta$, with $r$ and 
$\theta$ the usual radial and zenith angle spherical coordinates), $s$ is the 
distance in the midplane from the origin ($s= \sqrt{r^2 - z^2}$), $H_s$ is the 
disk scale height, and $\rho_0$ is the density at the reference midplane 
distance $s_0$.  All parameters describing the dependence of $H_s$ with $s$ 
(and thus the flaring of the disk) are the same as in Paper II.  
We note that this adopted disk flaring is very similar to that
found in the hydrodynamical simulations (see Figure 11 in Vorobyov \& Basu 
2010).  On the other hand, the adopted gas surface density profile 
$\Sigma \propto r^{-1}$ is somewhat shallower than the 
$\Sigma \propto r^{-1.5}$ usually seen in hydrodynamical simulations of 
gravitationally unstable disks in the embedded phase of star formation.  
The density profile is truncated at \rdisk\ and normalized so that the total 
mass matches 
\mdisk\ given by the simulations.  The inner radius of the disk is set equal 
to the dust destruction radius, calculated (assuming spherical, blackbody dust 
grains) as
\begin{equation}
R_{disk}^{in} = \sqrt{\frac{L_*}{16 \pi \sigma T_{dust}^4}} \qquad ,
\end{equation}
where $L_*$ is the protostellar luminosity (see below) and $T_{dust}$ is the 
dust destruction temperature (assumed to be 1500 K; e.g., Cieza et al.~2005).  

As in both Papers I and II, the total internal luminosity of the protostar and 
disk at each point in the collapse from core to star contains six components:  
(1) luminosity arising from accretion from the core directly onto the 
protostar, (2) luminosity arising from accretion from the core onto the disk, 
(3) luminosity arising from accretion from the disk onto the protostar, (4) 
disk ``mixing luminosity'' arising from luminosity released when newly 
accreted material mixes with existing disk material, (5) luminosity arising 
from the release of energy stored in differential rotation of the protostar, 
and (6) photosphere luminosity arising from gravitational contraction and 
deuterium burning.  The first five components are calculated following Adams 
\& Shu (1986); further details can be found in Papers I and II. 
In this work, however, the second and third components are calculated using 
direct input from the simulations, i.e., $\dot{M}_{\rm d}$ and $\dot{M}_{\rm s}$,
respectively.  

The sixth 
component, the photosphere luminosity arising from gravitational contraction 
and deuterium burning, follows the pre-main sequence tracks of D'Antona \& 
Mazzitelli (1994) with opacities from Alexander et al.~(1989).  In Papers I 
and II an offset of $10^5$ years was assumed between the onset of collapse and 
the zero time of these tracks (see also Myers et al.~1998). 
In this work, we take an improved approach by noting that the zero-time
of the D'Antona \& Mazzitelli tracks corresponds to the time when deuterium
burning begins in the protostellar interior.  
Hence, we make use of a simple stellar polytropic model  and assume that 
the zero time of the pre-main sequence tracks corresponds to the moment when 
the temperature in the stellar interior reaches $5\times10^5$~K. In order 
words, the time offset in our models is now the sum of two quantities: the 
time needed to form the first hydrostatic core 
from a pre-stellar cloud core, which may vary depending on the initial core 
mass, angular momentum
and density distributions, and the time needed to ignite
deuterium burning in the forming central protostar, which is usually 
$2\times10^4$~yr after the first hydrostatic core formation.  

Neither the time evolution of the protostellar radius nor that of the 
photosphere luminosity is included in our models in a self-consistent 
manner.  \rstar\ is 
calculated from \mstar\ following Palla \& Stahler (1991), who assume 
a constant accretion rate of 10$^{-5}$ \msun\ yr$^{-1}$.  The photosphere 
luminosity is included by adopting the pre-main sequence tracks of 
D'Antona \& Mazzitelli (1994).  These tracks do not include accretion and 
suffer from the uncertainty of how to define the zero time relative to the 
onset of collapse and formation of the protostar.  In reality, 
the accretion luminosity depends on the 
protostellar radius (\lacc\ $\propto$ 1/\rstar), and both the 
magnitude and zero time of the photosphere luminosity depend on the 
accretion history (Baraffe et al.~2009; Hosokawa et al.~2011; Hartmann 
et al.~2011).  Thus the accuracy of both components of the total model 
luminosity can be improved by including self-consistent calculations of the 
protostellar radius and photosphere luminosity.  Such calculations can now 
be performed for any arbitrary accretion history 
(e.g., Baraffe et al.~2009), and will be explored in a future paper.

Finally, there is also external luminosity arising from heating of the core by 
the Interstellar Radiation Field (ISRF).  We adopt the same ISRF as Papers I 
and II:  the Black (1994) ISRF, modified in the ultraviolet to reproduce the 
Draine (1978) ISRF, and then extincted by $A_V = 0.5$ of dust with properties 
given by Draine \& Lee (1984) to simulate extinction by the surrounding lower 
density environment.  We input the mean intensity of this ISRF ($J_{ext}$) 
into the radiative transfer code as an additional source of heating.  The 
luminosity added to \lbol\ from this external heating, $L_{ext}$, is 
determined after each radiative transfer model is run by subtracting the total 
internal luminosity (the sum of the above six components) from the total model 
luminosity.

For each model, we use the two-dimensional, axisymmetric, Monte Carlo dust 
radiative transfer package RADMC (Dullemond \& Turolla 2000; Dullemond \& 
Dominik 2004) to calculate the two-dimensional dust temperature profile of the 
core throughout the duration of each model.  Each model begins when the FHSC 
forms and terminates when 10\% of the initial core mass remains, which we have 
taken to be the Class I/II boundary (Vorobyov 2009b; Vorobyov \& Basu 2010).  
Some models with initially low core masses are extended until 
1\% of the initial core mass remains.  
The simulation output is resampled onto a timestep grid appropriate for the 
radiative transfer models; a detailed description of this resampling and its 
effects on the results are given in Appendix \ref{app_resamp}.  For the dust 
properties, we adopt the dust opacities of Ossenkopf \& Henning (1994) 
appropriate for thin ice mantles after $10^5$ yr of coagulation at a gas 
density of $10^6$ cm$^{-3}$ (OH5 dust), which have been shown to be 
appropriate for cold, dense cores (e.g., Evans et al.~2001; Shirley et 
al.~2005).  Isotropic scattering off dust grains is included in the model as 
described in Paper II.  Spectral Energy Distributions (SEDs) at each timestep 
are then calculated at 9 different inclinations ranging from $i=5-85$\degree\ 
in steps of 10\degree.  An inclination of $i=0$\degree\ corresponds to a 
pole-on (face-on) system, while an inclination of $i=90$\degree\ corresponds 
to an edge-on system.

\section{Comparing Models to Observations}\label{sec_comparing}

In this section we briefly summarize the method we use to turn the models 
into observational signatures and the observational dataset to which we 
compare the models.  Further details can be found in Paper II.

\subsection{Calculating Observational Signatures}\label{sec_evolsigs}

We use the model SEDs to calculate observational signatures of the models at 
each timestep for each inclination.  We calculate the bolometric luminosity 
(\lbol), the ratio of bolometric to submillimeter luminosity (\lbolsmm), and 
the bolometric temperature (\tbol).  \lbol\ is calculated by intergrating over 
the full SED,
\begin{equation}\label{eq_lbol}
\lbol = 4\pi d^2 \int_0^{\infty} S_{\nu}d\nu \qquad ,
\end{equation}
while the submillimeter luminosity is calculated by integrating over the SED 
for $\lambda$ $\geq$ 350 \um,
\begin{equation}\label{eq_lsmm}
\lsmm = 4\pi d^2 \int_0^{\nu=c/350\, \mu m} S_{\nu}d\nu \qquad .
\end{equation}
The bolometric temperature is defined to be the temperature of a blackbody 
with the same flux-weighted mean frequency as the source (Myers \& Ladd 
1993).  Following Myers \& Ladd, \tbol\ is calculated as
\begin{equation}\label{eq_tbol}
\tbol = 1.25 \times 10^{-11} \, \frac{\int_0^{\infty} \nu 
S_{\nu} d\nu}{\int_0^{\infty} S_{\nu} d\nu} \quad \rm{K} \qquad .
\end{equation}
\tbol\ can be thought of as a protostellar equivalent of $T_{eff}$; \tbol\ 
starts at very low values ($\sim 10$ K) for cold, starless cores and 
eventually increases to $T_{eff}$ once the core and disk have fully 
dissipated.  The integrals defined in equations \ref{eq_lbol} $-$ 
\ref{eq_tbol} are calculated using the trapezoid rule to integrate the 
finitely sampled model SEDs.

\subsection{Observational Dataset}\label{sec_obs}

We use the 1024 Young Stellar Objects (YSOs) in the five large, nearby 
molecular clouds surveyed by the \emph{Spitzer Space Telescope} Legacy 
Project ``From Molecular Cores to Planet Forming Disks'' (Evans et al.~2003) 
as our observational dataset.  Evans et al.~(2009) compiled photometry and 
calculated \lbol\ and \tbol\ for all 1024 YSOs in the same manner as described 
above after correcting the photometry for foreground extinction (see Evans et 
al.~[2009] for details).  They concluded that 112 of the 1024 YSOs are 
embedded protostars based on association with a millimeter continuum emission 
source tracing a core.  Since the models presented here are those of cores 
collapsing to form protostars, it is to these 112 objects that we compare.  
Core masses for all objects in Perseus, Ophiuchus, and Serpens are taken from 
Enoch et al.~(2009a).

\section{Results}\label{sec_results}

\begin{figure*}[hbt!]
\epsscale{1.0}
\plotone{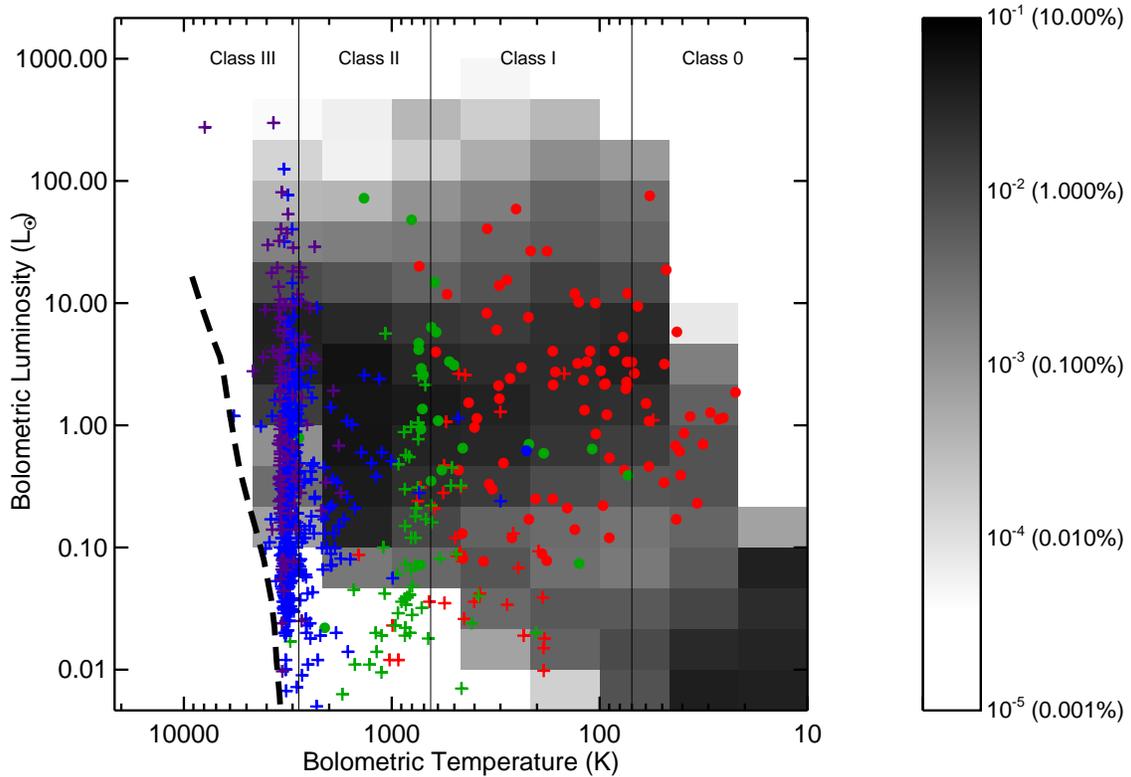}
\caption{\label{fig_blt}Combined BLT diagram for all models, weighted by mass 
and inclination as described in the text.  The grayscale pixels indicate the 
fraction of total time the models spend in each \lbol\ $-$ \tbol\ bin, 
calculated from Equation \ref{eq_bins}.  The grayscale is displayed in a 
logarithmic stretch with the mapping between grayscale and fraction of total 
time as indicated in the legend.  The class boundaries in \tbol\ are taken 
from Chen et al.~(1995).  The thick dashed line shows the ZAMS (D'Antona \& 
Mazzitelli 1994) from 0.1 to 2.0 \msun.  The colored symbols show the Young 
Stellar Objects from Evans et al.~(2009) in this diagram; the color indicates 
spectral class (red for Class 0/I, green for flat spectrum, blue for Class II, 
and purple for Class III), while a circle or cross indicates the source is or 
is not, respectively, associated with a core as traced by millimeter continuum 
emission.}
\end{figure*}

Figure \ref{fig_blt} compares the models considered here to observations by 
plotting \lbol\ vs.~\tbol.  Such a plot was first introduced by Myers et 
al.~(1998), who called it a BLT (bolometric-luminosity-temperature) diagram, 
as a protostellar equivalent to the Hertzsprung-Russell diagram.  Deeply 
embedded protostars surrounded by dense dust cores generally start at very low 
\lbol\ and \tbol\ and increase in both quantities as the dust core dissipates 
through accretion and mass-loss and the source gains mass (and thus 
luminosity).  \tbol\ eventually approaches \teff\ as the surrounding dust 
fully dissipates.  Indeed the stellar main sequence is plotted on Figure 
\ref{fig_blt}.  Also plotted are the 1024 YSOs from Evans et al.~(2009), with 
color indicating spectral class (red for Class 0/I, green for Flat spectrum, 
blue for Class II, and purple for Class III; see Evans et al.~for details) 
and symbol indicating source type (circles for sources associated with 
envelopes as traced by millimeter continuum emission, plus signs for sources 
not associated with envelopes).

To compare the models and observations, we determine the fraction of total 
time the models spend in various bins in the BLT diagram.  To calculate this 
fraction, we first divide the \lbol\ $-$ \tbol\ space into bins of 1/3 dex in 
both dimensions and then calculate the fraction of total time the models spend 
in each \lbol\ $-$ \tbol\ bin ($f_{bin}$) as

\begin{equation}\label{eq_bins}
f_{bin} = \frac{\displaystyle\sum_{mass} \left( \displaystyle\sum_{inc} 
t_{bin}w_{inc} \right) w_{mass}}{\displaystyle\sum_{mass} \left( 
\displaystyle\sum_{inc} t_{model}w_{inc} \right) w_{mass}} \qquad ,
\end{equation}
where the numerator is the time spent in the bin and the denominator is the 
total time.  The interior sum in both the numerator and denominator is over 
the 9 different inclinations while the exterior sum is over the different 
initial core masses.  The quantity $t_{bin}$ is the total time that a 
particular model viewed at a particular inclination spends in the specified 
\lbol\ $-$ \tbol\ bin whereas $t_{model}$ is the total duration of each 
individual model.  $w_{inc}$ is the weight each inclination receives in the 
sum, defined as the fraction of solid area subtended by that inclination.  As 
in Paper II, this is calculated in practice by assuming each of the 9 SEDs 
calculated is valid for inclinations spanning the range ($i-5$\degree) to 
(i+5\degree).

The final quantity in Equation \ref{eq_bins} is $w_{mass}$, the weight given 
to each of the individual models.  In Paper II we assigned these weights 
according to the initial core mass and the empirically derived core mass 
function (CMF) of starless cores (Motte et al.~1998; Testi \& Sargent 1998; 
Johnstone et al.~2000, 2001; Motte et al.~2001; Johnstone \& Bally 2006; 
Alves et al.~2007; Nutter \& Ward-Thompson 2007; Enoch et al.~2008; Hatchell 
\& Fuller 2008; Simpson et al.~2008; Rathborne et al.~2009; Sadavoy et 
al.~2010).  However, as noted in Paper II, there is considerable variation in 
the CMF between the various studies cited above, particularly below an initial 
core mass of $\sim$ 1 \msun\ since most of these studies feature completeness 
limits between $0.1-1$ \msun.  Thus, in this paper we instead assign the 
weights according to the final stellar mass produced by each model and the 
stellar initial mass function (IMF).  We adopt the Kroupa (2002) 
three-component power law IMF, which gives $dN/dM \propto M^{-\alpha}$, with 
$\alpha=0.3$ for $0.01 \leq M/\msun < 0.08$, $\alpha=1.3$ for $0.08 \leq 
M/\msun < 0.5$, and $\alpha=2.3$ for $M/\msun \geq 0.5$.  As noted by Kroupa 
(2002), the IMF may actually be better described by a four-component 
power-law, with $\alpha=2.7$ for $M/\msun \geq 1$ if uncertain corrections for 
unresolved multiple systems are applied.  As these corrections are very 
uncertain we do not adopt them here.

With the fraction of total time spent in each bin in the BLT diagram 
calculated as described above, we plot the results as grayscale pixels in 
Figure \ref{fig_blt}.  Since the models span only the duration of the embedded 
stage when the protostar and disk are still surrounded by the dense core from 
which they are forming, the relevant comparison is to the 112 embedded sources 
associated with cores as traced by millimeter continuum emission and thus 
plotted as filled circles.  The models clearly reproduce the full spread of 
observations of embedded sources in this diagram, unlike models with constant 
mass accretion, which only reproduce the high luminosity end of this 
spread (see, e.g., Figure 19 of Paper I 
and Figures 1, 7, 11, and 16 of Paper II).  While at first glance it appears 
the models overpredict the fraction of total time at high luminosities 
compared to the observations, this is an artifact of the logarithmic scaling.  
The models only spend 0.2\% of the time above 100 \lsun; in a dataset of 112 
protostars, there should be $<1$ object at such high luminosities, consistent 
with the observations.

\begin{figure}[hbt!]
\epsscale{1.0}
\plotone{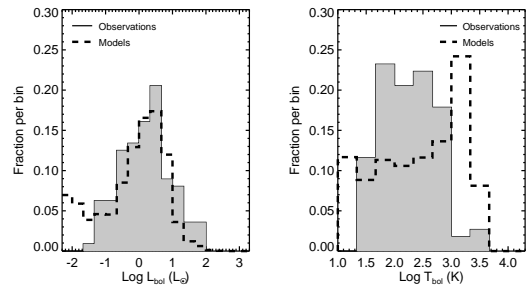}
\caption{\label{fig_lboltbolhist}Histograms showing the fraction of total 
sources (observations; solid filled histogram) and fraction of total time 
spent by all models (dashed unfilled histogram; calculated from Equation 
\ref{eq_bins}) at various \lbol\ (left) and \tbol\ (right).  The binsize is 
1/3 dex in both quantities.  For the observations, only the 112 embedded 
sources (plotted as filled circles on the BLT diagrams) are included.}
\end{figure}

We also compare the models to the observed \lbol\ and \tbol\ distributions 
separately in Figure \ref{fig_lboltbolhist}.  The observational histograms in 
Figure \ref{fig_lboltbolhist} only include the 112 embedded sources associated 
with envelopes (plotted as filled circles in Figure \ref{fig_blt}) and plot 
the fraction of total sources in each bin, while the model histograms plot the 
fraction of total time spent in each bin calculated from Equation 
\ref{eq_bins}.  The left panel of Figure \ref{fig_lboltbolhist} shows that the 
models generally reproduce the observed protostellar luminosity distribution; 
a K-S test on the observed and model \lbol\ distributions returns a value of 
0.85, indicating the two distributions are quite similar.  Most of the 
difference between the model and observed distributions actually exists in a 
``reverse luminosity problem,'' a point we will return to in \S 
\ref{sec_discussion_lum}.  The right panel suggests that the models do not 
provide as good a match to the observed \tbol\ distribution; a K-S test on the 
observed and model \tbol\ distributions returns a value of 0.42, confirming 
this observation.  Most of the difference is in a population of embedded 
objects at high \tbol\ ($\ga 1000$ K) predicted by the models but lacking in 
the observations.  We will discuss this discrepancy in detail in \S 
\ref{sec_discussion_tbol} below, but note here the main point that this 
discrepancy does not affect our conclusions regarding the match between 
observed and model luminosities.

\begin{figure*}[hbt!]
\epsscale{1.0}
\plotone{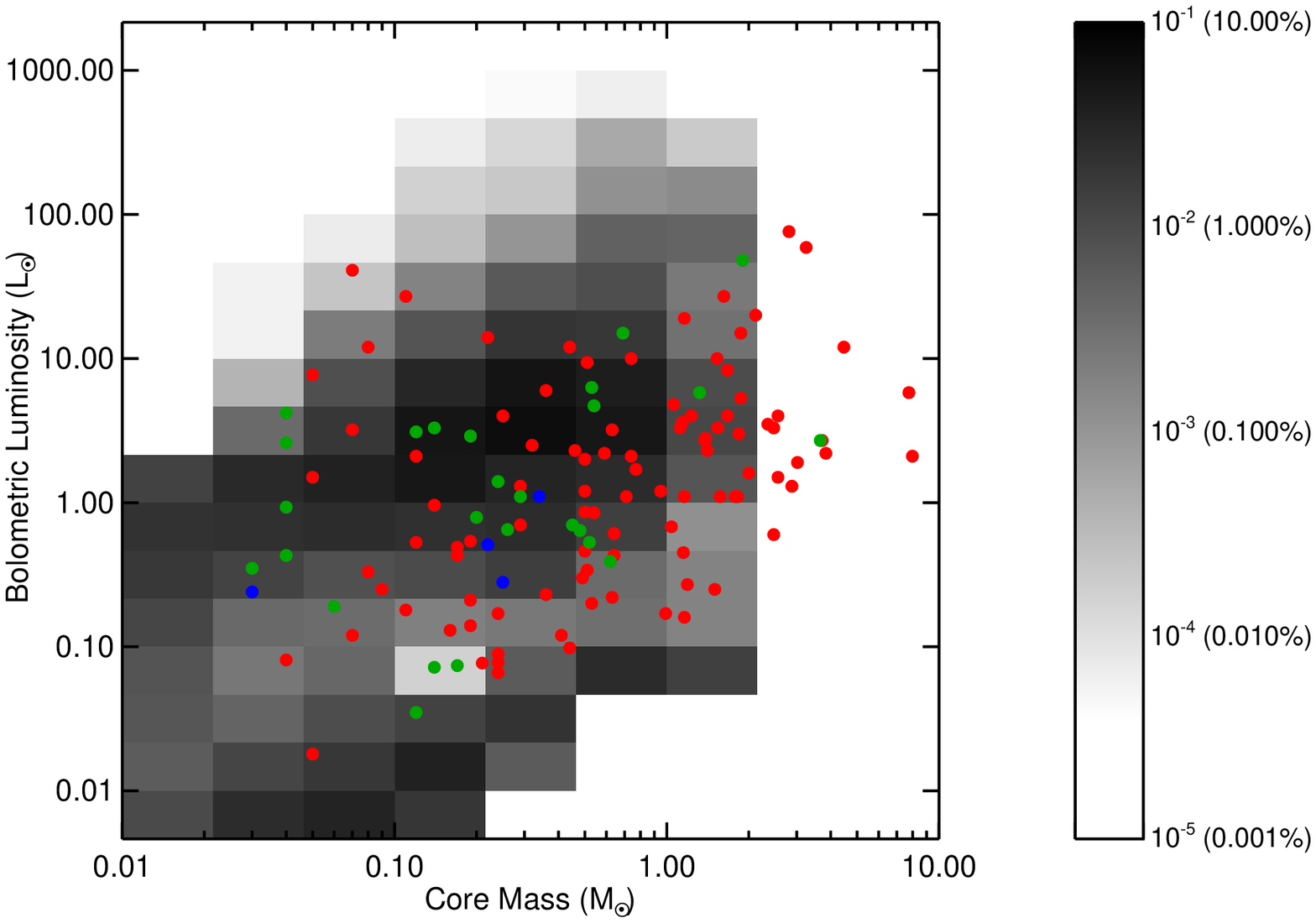}
\caption{\label{fig_lbolmenv}Combined plot of \lbol\ vs.~\menv\ for all 
models, weighted by mass and inclination as described in the text.  The 
grayscale pixels indicate the fraction of total time the models spend in each 
\lbol\ $-$ \menv\ bin, calculated from Equation \ref{eq_bins}.  The grayscale 
is displayed in a logarithmic stretch with the mapping between grayscale and 
fraction of total time as indicated in the legend.  The colored symbols show 
the Young Stellar Objects from Evans et al.~(2009) in this diagram, with 
\menv\ taken from Enoch et al.~(2009a) and the color used to indicate spectral 
class (red for Class 0/I, green for flat spectrum, blue for Class II, and 
purple for Class III).}
\end{figure*}

Figure \ref{fig_lbolmenv} compares the models and observations on a 
plot of \lbol\ vs.~\mcore.  The filled circles show the same embedded sources 
as in Figure \ref{fig_blt}, with color again indicating spectral class.  The 
core masses could easily shift by factors of $\sim 2-4$ in both directions 
depending on the true values of the dust opacity (e.g., Shirley et al.~2005; 
2011) and dust temperature in each core.  Furthermore, both \lbol\ and 
\mcore\ have variable completeness limits since the star-forming regions 
surveyed by c2d are not all located at the same distance.  Even for sources 
at the same distance the completeness limits are difficult to quantify with 
a single number.  For \lbol, the c2d survey is generally complete to 
protostars with \lbol\ $\ga$ 0.05 \lsun, although the exact completeness 
limit varies depending on the detailed shape of the SED of each source (Dunham 
et al.~2008; Evans et al.~2009; Enoch et al.~2009a).  For \mcore, the 
observations presented by Enoch et al.~(2009a) are generally complete above 
0.8 \msun, but the exact completeness limit depends on core size due to 
observing and data reduction limitations (see Enoch et al.~2008 for details).  
Thus, we simply note that the models do appear to reproduce the full spread of 
sources in \lbol\ $-$ \mcore\ space, including the existence of VeLLOs with 
\lbol\ $\sim$ $0.1-0.2$ \lsun\ and \mcore\ $\sim$ 1 \msun\footnote{As stated 
in \S \ref{sec_intro}, a VeLLO is, by definition, a protostar with \lint 
$\leq$ 0.1 \lsun, but external heating will raise \lbol\ to higher values.  
For example, the VeLLO IRAM04191$+$1522 has \lint\ $\sim$ 0.08 \lsun\ but 
\lbol\ $\sim$ 0.15 \lsun\ (\andre\ et al.~1999; Dunham et al.~2006).}.  This 
diagram should be revisited in the future once deeper submillimeter and 
millimeter continuum surveys of nearby star-forming regions (such as the 
upcoming SCUBA-2 Gould Belt survey; Ward-Thompson et al.~2007b) have been 
completed to evaluate whether or not the time spent at \lbol\ $\la$ 1 \lsun\ 
and \mcore\ $\la$ 0.1 \msun, as predicted by the models, is matched by 
observations of embedded sources.

\section{Discussion}\label{sec_discussion}

\subsection{Resolving the Luminosity Problem}\label{sec_discussion_lum}

As shown above in Figures \ref{fig_blt} and \ref{fig_lboltbolhist}, the models 
considered here reproduce the full range of observations in \lbol\ $-$ \tbol\ 
space and provide a reasonable match to the observed protostellar luminosity 
distribution.  Thus we conclude that the accretion process predicted by the 
Vorobyov \& (2005b, 2006, 2010) simulations of collapsing cores resolves 
the luminosity problem, although we caution that these results must be 
revisited in the future once simulations that feature smaller sink cells 
closer to the stellar surface and fully capture both the physics in the 
inner disk and the actual protostellar accretion rates are possible (see 
\S \ref{sec_simulations_accretion} for further discussion).  
In Paper II we argued 
that episodic accretion is both necessary and sufficient to resolve the 
luminosity problem, but that conclusion is subject to uncertainty since the 
actual prescription for episodic accretion included in that paper was quite 
simple and did not fully capture the accretion process in the simulations.  
Here we have coupled the exact evolution of the collapsing cores with 
radiative transfer calculations and demonstrated that the Vorobyov \& Basu 
simulations resolve the luminosity problem, although we defer the question of 
whether or not episodic accretion itself is necessary to Section 
\ref{sec_discussion_bursts} below.  We also note here that these models 
predict a smooth distribution in \lbol\ 
$-$ \tbol\ space, whereas the models in Paper II featured white ``excluded 
zones'' where the models spent no time but sources were observed to exist.  
The fact that only 3 initial mass cores were considered in Paper II (0.3, 1, 
and 3 \msun) was argued to artifically create these zones; the increased 
sampling of core masses in this paper confirms this argument.

Most of the remaining discrepancy between the model predictions and 
observations of protostellar luminosities actually appears in the form of a 
``reverse luminosity problem'', evident in both Figures \ref{fig_blt} and 
\ref{fig_lboltbolhist} as an overabundance of time spent at \lbol\ $\la$ 0.1 
\lsun\ compared to observations.  Indeed, a K-S test on the observed and model 
luminosity distributions that only compares the distributions above 0.1 \lsun\ 
returns a value of 0.87, slightly higher than the value of 0.85 obtained for 
the full distributions.  What causes the disagreement at low luminosities?  
Only 7\% of the observed 
sources have \lbol\ $\leq 0.1$ \lsun\ whereas the models spent 21\% of the 
total time at such luminosities.  Furthermore, the majority of this time (19\% 
out of 21\%) is spent at \tbol\ $\leq 100$ K, where there are no observed 
sources.  At least some, and possibly all, of this discrepancy can be 
explained by observational completeness limits.  As noted above, the c2d 
survey is generally complete to protostars with \lbol\ $\ga$ 0.05 \lsun, 
although the exact completeness limit varies depending on distance and the 
detailed shape of the SED of each source.  Indeed, several extremely low 
luminosity sources undetected in the \emph{Spitzer} c2d survey, with both 
internal and bolometric luminosities significantly below 0.1 \lsun, have 
recently been discovered, most through detections of outflows driven by cores 
with no associated \emph{Spitzer} c2d sources (Chen et al.~2010; Enoch et 
al.~2010; Dunham et al.~2011; Pineda et al.~2011).  With such low luminosities 
at least some of these sources may be first hydrostatic cores.  Sensitive 
interferometer outflow surveys and very deep \emph{Herschel} and \emph{James 
Webb Space Telescope} infrared surveys directed towards cores currently 
classified as starless are needed to fully identify and characterize the 
population of such extremely low luminosity protostars and/or first cores 
before an accurate comparison to the models can be made for luminosities 
below 0.1 \lsun.

\subsection{Bolometric Temperatures}\label{sec_discussion_tbol}

As noted above in \S \ref{sec_results}, the models do not provide 
a good match to the observed \tbol\ distribution, with most of the 
discrepancy in a population of embedded objects at high \tbol\ 
($\ga 1000$ K) predicted by the models but lacking in the observations.  
In the models, most of this time spent at \tbol\ $\ga 1000$ K arises 
when \rdisk\ is larger than a few hundred AU and the line-of-sight 
does not pass through the disk.  As described in detail in \S 
\ref{sec_model_radtrans}, we adopt analytic profiles for the disk and core 
density profiles since the hydro simulations do not provide the full vertical 
density structure.  In order to do this, we define the core inner radius to 
be equal to the disk outer radius so that there is no overlap between 
where the disk and core density profiles are defined.  However, as a 
consequence of this method, large cavities devoid of material exist above 
the surface of the disk but within the core inner radius, and these cavities  
increase as the disk sizes increase.  Lines of sight that pass through these 
cavities have reduced optical depths, allowing more short-wavelength 
emission to escape and thus increasing \tbol.

In reality, such large cavities are unlikely to exist; instead, the 
disk and core density profiles should smoothly join together.  As we have 
mentioned in \S \ref{sec_model_radtrans}, an improved methodology that 
reconstructs the disk and core vertical structure and incorporates this exact 
structure into the radiative transfer calculations (rather than adopting 
simple analytic profiles) is currently under development and will be presented 
in a future paper.  This method will result in a more accurate distribution 
of material above the disk surface and will likely remove much of the 
discrepancy between observed and model values of \tbol.  Since the 
distribution of luminosities is set mainly by the accretion rates and 
protostellar masses, we argue that including a more accurate physical 
structure should not significantly alter our results on the resolution of the 
luminosity problem, although this will be explicitly tested in a future 
paper.  We also note here the possibility that these models contain too much 
rotation and angular momentum, since this would cause both the rotational 
flattening of the cores and sizes of the disks to be overestimated, resulting 
in the models overpredicting \tbol\ compared to observations.  As discussed 
in some detail in \S \ref{sec_model_initial} above we do not consider this 
to be very likely, but we acknowledge that it is a possibility since $\beta$, 
the ratio of rotational to gravitational energy, is not a directly observable 
quantity and observed ranges of $\beta$ could be overestimated if infall 
and/or outflow motion is mistakenly attributed to rotation.

\begin{figure}
\epsscale{1.0}
\plotone{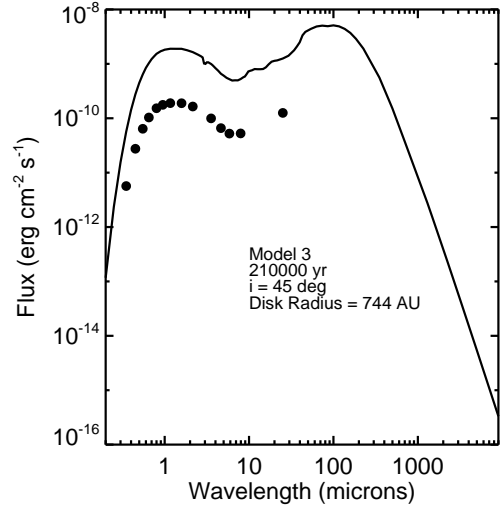}
\caption{\label{fig_tbolsed}Example of a Spectral Energy 
Distribution (SED) for a model with high \tbol.  This particular SED 
is for the $i=$ 45$^{\rm o}$ line-of-sight at 210,000 yr into the 
collapse of model 3, when the core mass is 0.9649 \msun, the disk mass is 
0.1656 \msun, the protostellar mass is 0.5122 \msun, and the disk radius is 
744 AU.  This SED has \tbol\ $= 1037$ K.  
The solid line plots the model SED over all wavelengths, while 
the points show the model SED at standard optical and near-infrared 
wavelengths and the $3.6-70$ \um\ wavelengths provided by the 
\emph{Spitzer Space Telescope}.  The points have been shifted down by a 
factor of 10 for clarity.}
\end{figure}

Even after including a more accurate physical structure, some 
discrepancy between observed and model values of \tbol\ may remain, 
particularly once the effects of outflow cavities are included.  Figure 
\ref{fig_tbolsed}, which plots an example spectral energy distribution (SED) 
for a model with high \tbol, shows that such models feature a double-peaked 
SED.  This figure also demonstrates that, from optical wavelengths 
to either the 24 or 70 \um\ bands probed by \emph{Spitzer} observations, 
such models resemble transition disk Class II sources (sometimes also 
referred to as cold disk sources; see \merin\ et al.~[2010] and references 
therein).  Given that our models predict that 40\% of embedded sources 
are classified as Class II by \tbol, and that the observed fraction of 
Class 0+I to Class II sources is 0.19 when classifying via 
extinction-corrected values of \tbol\ (Evans et al.~2009), our models 
predict that 12\% of Class II young stellar objects are actually embedded 
objects with SEDs like that shown in Figure \ref{fig_tbolsed}.  While this is 
consistent with the upper range of the observed fraction of Class II sources 
with transition disk SEDs (3\% -- 12\%; \merin\ et al.~2010; Furlan et 
al.~2011), the models 
clearly overpredict this fraction, as described above.  Nevertheless, we 
caution that a small fraction of Class II sources with transition disk 
SEDs could in fact be embedded sources.

If the above statement is true, why are they missing from the 
observed sample of embedded sources?  As discussed in detail in Paper II, 
whether or not a population of embedded objects with high \tbol\ exists 
remains an open question.  The Evans et al.~(2009) sample is based primarily 
on the association of \emph{Spitzer} sources featuring rising SEDs and red 
colors with millimeter continuum sources (Enoch et al.~2009a; Dunham et 
al.~2008) and is likely biased against such objects since they would often not 
meet these criteria and would instead be assumed to be chance alignments 
between millimeter continuum sources and background sources and/or later-stage 
YSOs.  Furthermore, the 
extinction corrections applied by Evans et al.~(2009) to the embedded sources 
are average extinctions over each individual cloud rather than true 
line-of-sight extinctions, and thus may underestimate the true extinction 
since current, active star formation (and thus the position of the youngest, 
embedded sources) is associated with the densest parts of molecular clouds 
(e.g., Heiderman et al.~2010; Lada et al.~2010).  As shown in Paper II, such 
underestimates could, in the worst cases, artifically lower \tbol\ from 
several thousand K to several hundred K.  Future work must revisit the 
observational samples and carefully evaluate whether or not a population of 
embedded sources with high enough \tbol\ to be classified as Class II or Class 
III exists.

\subsection{Duration of the Embedded Phase}\label{sec_discussion_durations}

\begin{figure}
\epsscale{1.0}
\plotone{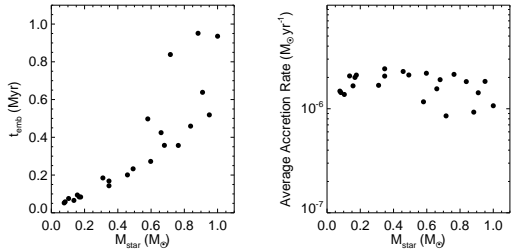}
\caption{\label{fig_duration}\emph{Left:} $t_{emb}$ vs.~\mstar\ for the 23 
models listed in Table \ref{tab_models} and considered in this paper.  
\emph{Right:}  Average \mdotstar\ over the embedded phase duration 
vs.~\mstar\ for the same models.}
\end{figure}

For the 23 models listed in Table \ref{tab_models} and considered in this 
paper, the duration of the embedded phase ($t_{emb}$) ranges from 0.052 
(model 19) to 0.951 Myr (model 11).  For each model, 
Figure \ref{fig_duration} plots both 
$t_{emb}$ (left panel) and the average accretion rate over the embedded phase 
duration (right panel, calculated as final stellar mass divided by $t_{emb}$) 
vs.~the final stellar mass produced by the model.  The duration of the 
embedded phase increases approximately linearly with the final stellar 
mass produced, and thus the average mass accretion rate (not to be confused 
with the \emph{instantaneous} mass accretion rate, which is highly variable 
[see Figure \ref{fig_mdot}]) is approximately constant and does not depend on 
final stellar mass.  Offner \& McKee (2011) argued that 
models that tend toward constant accretion time rather than constant accretion 
rate are necessary to match the observed protostellar luminosity 
distribution, consistent with the conclusion by Myers (2010) that 
accretion rates that increase with mass can at least partially resolve the 
luminosity problem.  We disagree that such models 
are necessary.  Our models tend toward a constant average accretion rate 
of $\sim 1-3 \times 10^{-6}$ \msun\ yr$^{-1}$ rather than a constant 
accretion time yet still provide an excellent match to the observed 
protostellar luminosity distribution.

The average duration of the embedded phase, weighted by final stellar mass as described in 
\S \ref{sec_results}, is 0.12 Myr.  In contrast, Evans et al.~(2009) derived 
an embedded phase lifetime of $t_{emb} = 0.44$ Myr based on the ratio of 
Class 0+I sources to Class II sources in the c2d sample and the assumption of 
a Class II lifetime of 2 Myr.  Thus these models predict a significantly 
shorter $t_{emb}$ than suggested by recent observations.  However, a number of 
caveats apply to this comparison:  (1) We have taken the Class I/II boundary 
to be the point at which 10\% of the initial core mass remains 
and terminate the models at this point.  In 
reality the exact point at which to set this Class boundary is uncertain and 
could easily shift the duration of the embedded phase by factors of $\sim$ 2 in either 
direction; (2) The observationally determined $t_{emb}$ is pinned to a Class 
II lifetime of 2 Myr, but the uncertainty in this lifetime is about 1 Myr 
(50\%; see discussion in Evans et al.~2009); and (3) The number of 
Class 0+I sources in the Evans et al.~sample may be overestimated.

The third point above is emphasized by three recent studies.  
First, van Kempen et al.~(2009) observed 22 Class I sources in Ophiuchus in 
\hcop\ $J = 4-3$, a tracer of warm and dense gas, and showed that 11 (50\%) 
are not detected and thus show no evidence of being surrounded by a dense 
core. Second, McClure et al.~(2010) used their revised extinction law 
and classification method to show that greater than 50\% of the Class I sources 
in Ophiuchus are highly extincted disk sources no longer embedded within 
cores.  Third, Heiderman et al.~(2010) observed 53 Class I 
sources in a variety of nearby star-forming regions in \hcop\ $J = 3-2$, a 
tracer of dense gas, and showed that 31 (58\%) are not detected and thus not 
associated with a dense core.  All three studies likely represent upper 
limits to the true fraction of ``fake'' Class I sources.  In the cases of van 
Kempen et al.~(2009) and McClure et al.~(2010) 
this is because there is substantial evidence that 
Ophiuchus is located behind an extinction screen that, if not properly 
accounted for, will redden source SEDs and artifically increase the number 
of Class I sources (see, e.g., Figure 12 of Evans et al.~2009).  In the case 
of Heiderman et al.~(2010) this is because their study specifically targeted 
suspiscious Class I sources located in low extinction regions of clouds.  
We thus consider 50\% as an upper limit to the fraction of 
Class I sources in the Evans et al.~(2009) sample that are actually 
misclassified Class II sources.  Shifting 50\% of the Class I sources to 
Class II would decrease the Evans et al.~(2009) observationally determined 
value of $t_{emb}$ from 0.44 Myr to 0.2 Myr.  Combined with the other caveats 
mentioned above, the duration of the embedded phase predicted by our models 
is marginally consistent with observations but should be revisited in the 
future as uncertainties in the observations are improved.

Finally, the Stage 0 ([\mstar+\mdisk] / [\mstar+\mdisk+\mcore] $\leq 0.5$; 
\andre\ et al.~1993) 
duration for the 23 models considered in this paper ranges from 0.009 (model 
19) to 0.256 (model 15) Myr, with an average (again weighted by final stellar 
mass as described in \S \ref{sec_discussion}) of 0.027 Myr.  Compared to the 
total embedded duration of 0.12 Myr, our models predict that the Stage 0 
phase is only 23\% of the total embedded duration.  This is a natural 
consequence of the fact that these models feature average mass accretion 
rates that decrease with time (Vorobyov \& Basu 2010; see Figure 
\ref{fig_mdot}), thus the first 50\% of the mass 
will accrete from the core faster than the second 50\% of the mass.  Our 
results are generally consistent with recent observationally determined 
estimates of the lifetime of Stage 0 relative to the total embedded phase.  
For example, Enoch et al.~(2009a) found 39 Class 0 
and 89 Class I sources in Perseus, Ophiuchus, and Serpens, giving a relative 
Stage 0 lifetime of 30\% of the total embedded phase duration.  
Additionally, Maury et al.~(2011) found that between $9-12$ of the 57 
protostars they identified in the Serpens South cluster 
(Gutermuth et al.~2008) were Class 0 sources, giving a relative Stage 0 
lifetime of 16\%$-$21\% of the total embedded phase duration.  While the 
Enoch et al.~and Maury et al.~results are difficult to compare quantitatively 
since they 
use different classification methods that trace the underlying physical 
Stage to different degrees of accuracy (\tbol\ in the case of Enoch et al.~and 
position in \lbol$-$\mcore\ space in the case of Maury et al.), the general 
agreement between our models and these observational results is encouraging.

\subsection{Number, Duration, and Importance of Bursts}
\label{sec_discussion_bursts}

The 23 models considered in this paper feature between 0 (model 16) and 97 
(model 15) accretion bursts, where the exact criteria for defining bursts is 
given in Appendix \ref{app_resamp}.  The percentage of total time spent in 
bursts ranges from 0\% (model 16) to 11.8\% (model 15), and the percentage of 
total mass accreted in bursts ranges from 0\% (model 16) to 35.5\% (model 4).  
On average (where the average is weighted by final stellar mass as described 
in \S \ref{sec_results}), 1.3\% of the total time is spent in bursts and 
5.3\% of the total mass is accreted in these bursts.  These values represent 
the statistical average values of the fraction of total time spent and mass 
accreted in bursts assuming a standard Kroupa IMF (see Section 
\ref{sec_results} for details).  We caution that the exact values 
depend on the criteria used for defining bursts and would increase if a lower 
accretion rate floor were used (see Appendix \ref{app_resamp}).

The simple, parameterized models presented in Paper II spend between 
$1.5\%-2$\% of their total time in bursts, consistent with the 1.3\% of total 
time featured by these models.  However, in the Paper II models between 
$50\%-91\%$ of the final stellar mass accretes in bursts, in stark contrast to 
the 5.3\% of total mass that accretes in bursts in the models considered 
here.  The explanation for this large difference lies in the detailed 
implementation of accretion bursts in Paper II.  As described in \S 
\ref{sec_intro}, a burst was triggered each time the ratio of \mdisk\ to 
\mstar\ exceeded 0.2.  At this point the accretion rate onto the star was 
increased from 0 \msun\ yr$^{-1}$ to $10^{-4}$ \msun\ yr$^{-1}$ until the disk 
was fully drained of mass, allowing the cycle to begin anew.  However, in 
reality only the most extreme bursts reach accretion rates of $10^{-4}$ 
\msun\ yr$^{-1}$, as evident from Figure \ref{fig_mdot}.  
Furthermore, only a very small amount of the mass in the disk accretes in a 
given burst (about $10^{-2}$ \msun, on average), whereas in Paper II the 
assumption that the disk was fully drained in each burst led to situations 
where up to $0.1-0.2$ \msun\ were accreting in single bursts.  The models 
considered here, based on actual hydrodynamic simulations rather than simple 
parameterizations, are significantly more realistic.

Comparison to the results presented in Papers I and II demonstrate 
that the accretion process predicted by the Vorobyov \& Basu (2005b, 2006, 
2010) simulations essentially resolves the luminosity problem inherent in 
models with constant mass accretion.  As first noted by Kenyon et al.~(1990), 
there are two types of non-steady mass accretion that could potentially 
resolve the luminosity problem: (1) accretion rates that start high and then 
decrease with time, and (2) generally low accretion rates punctuated by short, 
episodic bursts of high accretion.  Figure \ref{fig_mdot} clearly illustrates 
that the Vorobyov \& Basu simulations feature both declining accretion 
rates with time \emph{and} short, episodic accretion bursts.  Which of these 
is responsible for resolving the luminosity problem?

\begin{figure}
\epsscale{1.0}
\plotone{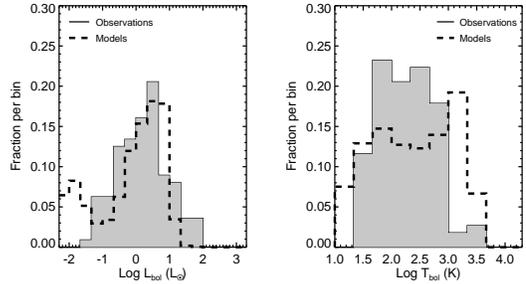}
\caption{\label{fig_lboltbolhist_avg}Histograms showing the fraction of total 
sources (observations; solid filled histogram) and fraction of total time 
spent by all models after time-averaging as described in \S 
\ref{sec_discussion_bursts} (dashed unfilled histogram; calculated from 
Equation \ref{eq_bins}) at various \lbol\ (left) and \tbol\ (right).  The 
binsize is 1/3 dex in both quantities.  For the observations, only the 112 
embedded sources (plotted as filled circles on the BLT diagrams) are included.}
\end{figure}

Given that, on average, only 5.3\% of the total mass accretes in 
bursts, one might suspect that 
it is the declining accretion rates with time rather than the bursts 
that lower model luminosities and improve the match to observations.  However, 
this 5.3\% excludes all of the mass that accretes in lower-amplitude 
accretion rate increases that do not meet the criteria for a burst 
as defined in \S \ref{app_resamp}.  Thus, to properly evaluate 
whether the bursts are required in order to resolve the luminosity problem, 
we have time-averaged the accretion rates to filter out the effects of the 
bursts and variability and re-run all 23 models.  
We averaged all models over 20,000 yr 
durations unless the total model duration was less than 0.2 Myr, in which case 
we decreased the averaging window to either 10,000 or 5,000 yr in order to 
preserve at least 10 timesteps between the formation of the protostar and the 
end of the embedded phase.

Figure \ref{fig_lboltbolhist_avg} plots histograms comparing the 
fraction of total time the time-averaged models spend at various \lbol\ and 
\tbol\ to the observed distributions, similar to Figure 
\ref{fig_lboltbolhist}.  While the time-averaged models provide a much better 
match to observations than models with constant mass accretion (see Papers 
I and II for such models), comparing Figures \ref{fig_lboltbolhist} and 
\ref{fig_lboltbolhist_avg} shows that the time-averaged models feature a 
small shift to higher luminosities and do not do quite as good of a job 
resolving the luminosity problem.  This is confirmed by a K-S test on the 
observed and time-averaged model luminosity distributions, which returns a 
value of 0.59, lower than the value of 0.85 returned for the original 
models.  

\begin{figure}
\epsscale{1.0}
\plotone{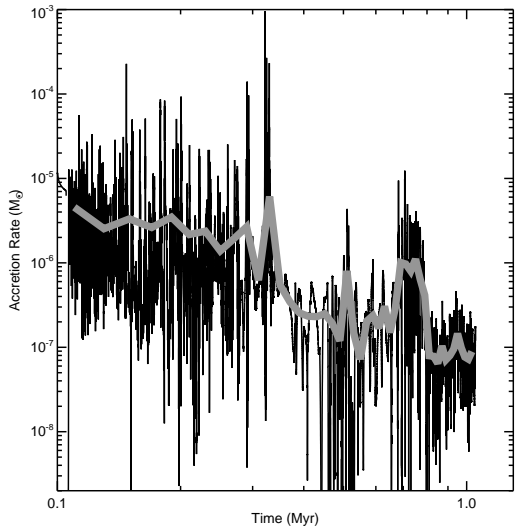}
\caption{\label{fig_arate_avg}\mdotstar\ vs.~time for the original (thin 
black line) and time-averaged (thick gray line) timesteps for Model 11.  
A window of 20,000 yr was used for the time-averaging.  Note that the 
variability and bursts have not been fully filtered out.}
\end{figure}

These results indicate that the declining accretion rates are not 
solely responsible for resolving the luminosity problem, a finding that 
is only reinforced by the fact that, even with the large windows over which we 
have time-averaged, the variability and bursts are not fully filtered 
out\footnote{In theory we could adopt even larger averaging windows, 
but to do so would decrease the total number of timesteps below 10 for many 
models and risk not fully sampling the decline of the accretion rate with 
time.} (see Figure \ref{fig_arate_avg} for an example).  We thus conclude 
that it is a combination of both the accretion rates that decline with time 
and the variability and episodic bursts that resolve the luminosity problem.  
We consider this to be a plausible result given that the Vorobyov \& Basu 
simulations self-consistently predict both, and we argue that the fact that 
the bursts are required is in general agreement with 
the other indirect evidence for accretion variability and bursts described 
in \S \ref{sec_intro}.

\section{Summary}\label{sec_summary}

In this paper we have coupled radiative transfer models with hydrodynamical 
simulations of collapsing cores predicting accretion rates that both decline 
with time and feature episodic accretion bursts caused by fragments torqued 
onto the protostar.  We have calculated the time evolution of standard 
observational signatures (\lbol, \tbol, and \lbolsmm) for cores collapsing 
following these simulations.  We have compared our results to a database of 
1024 YSOs containing 112 embedded protostars recently compiled by Evans et 
al.~(2009).  We summarize our main conclusions as follows:

\begin{enumerate}
\item The hydrodynamical simulations presented by Vorobyov \& Basu (2005, 
2006, 2010) reproduce the full spread of observed embedded protostars 
in a diagram of \lbol\ vs.~\tbol.  The models resolve the luminosity 
problem and provide a reasonable match 
to the observed protostellar luminosity distribution (K-S value of 0.85).  
The models predict a large number of sources at very low ($\la 0.1$ \lsun) 
luminosities absent in the observations due to the observational sensitivity 
limit; removing such low luminosities from the comparison between models and 
observations increases the K-S value to 0.87.  The models predict that only 
0.2\% of the total time is spent at \lbol\ $\geq$ 100 \lsun; a larger dataset 
than the 112 protostars considered here is necessary to test 
this prediction.  
Time-averaged models that filter out the accretion variability and 
 bursts do not provide as good of a match to the observed luminosity problem, 
suggesting that the bursts are required.
\item The models do not provide a good match to the distribution of observed 
\tbol\ for embedded protostars (K-S value of 0.42).  Instead, the models 
predict a subtantial population of embedded protostars at \tbol\ $\ga$ 1000 K 
and thus classified as Class II or Class III sources.  Most of this 
discrepancy arises from the method by which we adopted analytic disk and 
core density profiles and will be alleviated with future models that 
incorporate the exact physical structure from the hydro simulations, but some 
of the discrepancy may also be due to a population of embedded protostars 
with high values of \tbol\ missing from the current database of protostars 
due to the selection criteria applied to construct this database.  The 
planned future models will not significantly alter the model luminosity 
distribution since this is primarily determined by the accretion rates and 
protostellar masses, not by the detailed physical structure adopted in the 
radiative transfer calculations.
\item The models reproduce the full spread of sources in a plot of \lbol\ 
vs.~\mcore, including the existence of very low luminosity protostars with 
\lbol\ $\sim 0.1-0.2$ \lsun\ but relatively high core masses of $1-2$ \msun.
\item The duration of each model is approximately proportional to the final 
stellar mass produced, yet these models resolve the luminosity problem and 
provide an excellent match to the observed protostellar luminosity 
distribution.  This is in contrast to recent results in the literature 
claiming that models that tend toward constant accretion 
time rather than constant accretion rate are necessary to match the observed 
distribution.
\item The IMF-weighted average duration of the embedded phase in our models 
is 0.12 Myr, whereas Evans 
et al.~(2009) recently determined the embedded phase duration to be 0.44 
Myr.  We have suggested a number of possible means by which these two 
estimates of the embedded phase duration may be reconciled.  The IMF-weighted 
average model Stage 0 duration is 0.027 Myr, or 23\% of the total embedded 
phase duration.  Observationally determined values based on the ratio of Class 
0 to Class I sources range from $16\%-30\%$ (Enoch et al.~2009a; 
Maury et al.~2011).  Our models are consistent with this range.
\item On average, these models spend 1.3\% of their total time in accretion 
bursts, during which time 5.3\% of the final stellar mass accretes.  In the 
most extreme models these values reach 11.8\% and 35.5\%, respectively.  
Thus accretion is not truly ``episodic'' since it actually occurs at all 
times rather than only in episodes.  A better description is that accretion 
is ``variable with episodic bursts.''
\end{enumerate}

Future work must concentrate on improving the accuracy and self-consistency 
of the hydrodynamical simulations and radiative transfer models, and on 
compiling a more complete and more accurate database of protostars and their 
observational signatures.  Nevertheless, we expect our primary conclusion 
that the Vorobyov \& Basu (2005, 2006, 2010) simulations resolve the luminosity 
problem and match the observed protostellar luminosity distribution to 
remain unchanged to such future improvements.

\acknowledgments
We thank the referee for a careful reading of this paper and several comments 
that have significantly improved the quality of this work.  We also thank 
Neal Evans, Stella Offner, and Chris McKee for reading a draft of this 
paper in advance of publication and providing helpful comments and questions.  
Support for this work was provided by JPL contract 1433171 and RFBR grants 
10-02-00278 and 11-02-92601.   M.~M.~Dunham acknowledges travel support from 
an International Travel Grant from the AAS, and E.~I.~Vorobyov acknowledges 
support from a Lise Meitner fellowship.  The authors gratefully acknowledge 
N.~J.~Evans at The University of Texas at Austin for hosting both M.~M.~Dunham 
and E.~I.~Vorobyov at UT Austin to begin work on this project.  Support
for this visit was provided by NSF grant AST-0607793 to N.~J.~Evans and the UT 
Austin Tinsley Visitor's Fund.  This research has made use of NASA's 
Astrophysics Data System Bibliographic Services.

\appendix
\section{Resampling Timesteps}\label{app_resamp}

The simulations described in \S \ref{sec_model_sim} are calculated with 
data output timesteps of 20 yr (the physical timestep is even much smaller, 
of the order of several weeks), thus the time evolution of all quantities given by the 
simulations are output on grids with $2617-47554$ points for the models 
considered in this paper, which range in duration from $52340-951080$ yr (\S 
\ref{sec_discussion_durations}).  However, due to technical limitations, the 
radiative transfer models are limited to $\sim 500-600$ timesteps per model 
in order to be run in a reasonable amount of time.  Therefore we must resample 
the simulation output onto much coarser timestep grids.  As the primary 
motivation of this study is to evaluate the ability of the Vorobyov \& Basu 
(2010) simulations to resolve the luminosity problem, and the accretion 
luminosity is dominated by \mdotstar\ since \rstar\ is several orders of 
magnitude smaller than \rdisk\ for all but the earliest times, care must be 
taken to avoid altering the \mdotstar\ distribution.

Our resampling procedure is designed to alter the \mdotstar\ distribution as 
little as possible by ensuring that all accretion bursts are included in the 
resampled timesteps.  For each model, we define a floor in \mdotstar\ that 
starts a factor of 3 higher than \mdotstar\ at the moment of the formation of a protostar 
and declines at approximately the 
same average slope as \mdotstar\ throughout the duration of the model.  We 
define any duration of time where \mdotstar\ rises above this floor as a burst 
and combine bursts separated by less than 100 yr into a single burst.  After 
combining closely spaced bursts we are left with between $0-97$ bursts (\S 
\ref{sec_discussion_bursts}).  We then edit the burst timesteps by hand so 
that each burst occurs over no more than 5 timesteps, being careful to retain 
the first and last timesteps and also the timestep with the maximum \mdotstar\ 
so that the total duration and amplitude of each burst is preserved.  Finally, 
we sample the original timesteps every $N$ years, where $N$ ranges from 
$500-4000$ for the different models and is chosen such that the final number 
of resampled timesteps does not exceed $\sim 500-600$.  The final timestep 
grid is then constructed by combining the burst timesteps and regularly 
sampled timesteps, eliminating duplicate timesteps included both from the 
bursts and the regular sampling.

\begin{figure}
\epsscale{0.75}
\plotone{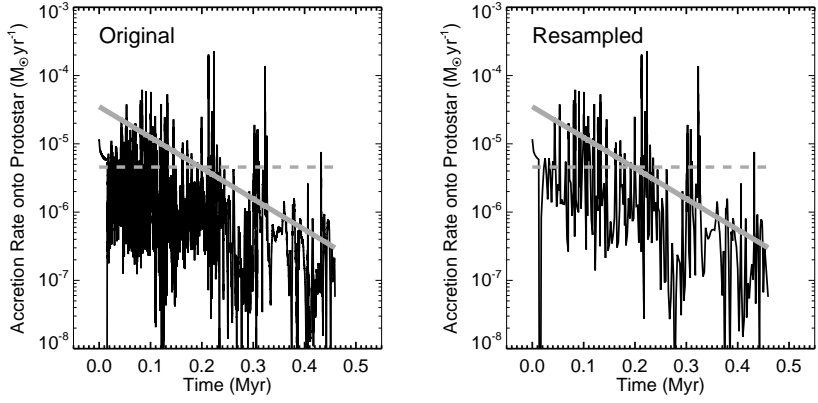}
\caption{\label{fig_resampexample}\mdotstar\ vs.~time for the original (left) 
and resampled (right) timesteps for Model 3.  In each panel the black line 
plots the \mdotstar\ given by the simulation, the solid gray line plots the 
\mdotstar\ floor used to define bursts, and the dashed gray line plots, for 
comparison, the constant accretion rate from the core onto the protostar+disk 
system of $4.57 \times 10^{-6}$ \msun\ yr$^{-1}$ from Paper II.}
\end{figure}

Figure \ref{fig_resampexample} shows an example of the resampling for Model 
3.  The left panel of this figure plots in black \mdotstar\ vs.~time for the 
original timesteps given by the simulation, whereas the right panel plots 
\mdotstar\ vs.~time after resampling the timesteps following the procedure 
described above.  Also plotted in each panel are the \mdotstar\ floor used to 
define bursts (solid gray line) and, for comparison, the constant accretion 
rate from the core onto the protostar+disk system of $4.57 \times 10^{-6}$ 
\msun\ yr$^{-1}$ from Paper II (dashed gray line).  Comparing the two panels 
shows that both the number and amplitude of the accretion bursts are 
preserved, as is the general decline in \mdotstar\ with time.

\begin{figure}
\epsscale{0.5}
\plotone{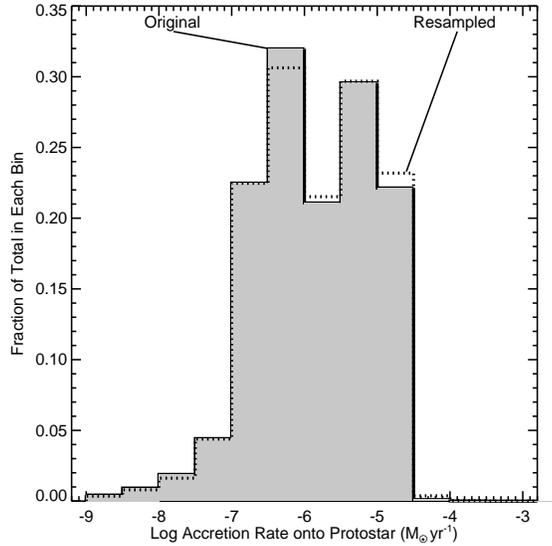}
\caption{\label{fig_mdothist}Histograms showing the fraction of total time 
spent in various \mdotstar\ bins for the original (solid, shaded histogram) 
and resampled (dashed histogram) timesteps.  The binsize is 1/2 dex 
in \mdotstar.}
\end{figure}

Careful inspection of Figure \ref{fig_resampexample} shows that not all 
features of the evolution of \mdotstar\ with time are preserved.  In 
particular, the episodes of lowest \mdotstar\ are not always preserved by the 
resampling procedure, with the most prominent case occuring at $\sim 0.12$ Myr 
in the above example.  This is a natural result of our adopted procedure since 
no special effort is devoted to preserving these episodes, thus they will only 
appear in the resampled timesteps if they happen to occur at a timestep 
included by the regular sampling.

To examine the overall effects of resampling on our results, Figure 
\ref{fig_mdothist} plots histograms of the fraction of total time the models 
spend in various \mdotstar\ bins for the original (solid, shaded histogram) 
and resampled (dashed histogram) timesteps.  The fraction of total time spent in each bin is calculated by 
dividing the total time spent in each bin (weighted by mass as described above 
in \S \ref{sec_results}) by the total duration of the models (again weighted 
by mass), with the equation

\begin{equation}\label{eq_bins_mdot}
f_{bin} = \frac{\displaystyle\sum_{mass} t_{bin} 
w_{mass}}{\displaystyle\sum_{mass} t_{collapse} w_{mass}} \qquad ,
\end{equation}
where $t_{bin}$ is the total time each model spends in the bin, $t_{collapse}$ 
is the total duration of each model, and $w_{mass}$ is the weight for each 
model depending on the final stellar mass produced (see \S 
\ref{sec_results}).  Figure \ref{fig_mdothist} shows that the original 
and resampled \mdotstar\ distributions are quite similar.  
There is a very small shift to higher values of 
\mdotstar\ after resampling seen in a careful inspection of the figure.  The 
models spend 0.5\% less time at \mdotstar\ $\leq 10^{-7}$ \msun\ yr$^{-1}$ and 
1.3\% more time at \mdotstar\ $\geq 10^{-5}$ \msun\ yr$^{-1}$ after 
resampling.  This small shift is a result of missing some of the lowest 
episodes of \mdotstar\ and has a negligible impact on the final luminosity 
distribution presented in \S \ref{sec_results} since timesteps with extremely 
low values of \mdotstar\ will have their total luminosity dominated by 
photosphere and external luminosity anyway.  We thus conclude that resampling 
to coarser timestep grids, as required by the radiative transfer models, does 
not fundamentally alter or bias any of the conclusions of this paper.

\end{document}